\documentclass[11pt]{article}
\pdfoutput=1

\usepackage{euscript}
\usepackage{amssymb}
\usepackage{amsfonts}
\usepackage{amsbsy}
\usepackage{epsfig}
\usepackage{amsthm}
\usepackage{amscd}
\usepackage{amstext}
\usepackage{verbatim}
\usepackage{amsmath}
\usepackage{cancel}
\usepackage{capt-of}
\usepackage{mathtools}

\def\ben{\begin{equation}}
\def\een{\end{equation}}
\def\half{{\textstyle{\frac12}}}

\let\a=\alpha \let\b=\beta \let\g=\gamma \let\d=\delta 
\let\z=\zeta   \let\k=\kappa
     
\let\s=\sigma

 \let\G=\Gamma   
   
\let\C=\Chi

\let\pa=\partial

\newcommand{\ba}{\begin{align}}
\newcommand{\ea}{\end{align}}

\newcommand{\bi}{\begin{itemize}}
\newcommand{\ei}{\end{itemize}}

\def\be{\begin{equation}}
\def\ee{\end{equation}}
\def\beq{\begin{equation}}
\def\eeq{\end{equation}}

\def\dalemb#1#2{{\vbox{\hrule height .#2pt
       \hbox{\vrule width.#2pt height#1pt \kern#1pt
               \vrule width.#2pt}
       \hrule height.#2pt}}}

\newcommand{\bea}{\begin{eqnarray}}
\newcommand{\eea}{\end{eqnarray}}
\newcommand{\tr}{{\rm tr} }

\def\vep{{\varepsilon}}

\def\C{{{\Bbb C}}}

\def\Z{{{\Bbb Z}}}

\def\ocal{{\mathcal{O}}}

\def\({\left(} \def\){\right)}
\def\[{\left[} \def\]{\right]}

\def\Real{{\rm Re}\,}

\begin{document}

\begin{flushright}
\end{flushright}

\begin{center}

{ \Large {\bf Holographic entanglement beyond classical gravity
}}

\vspace{1cm}

Taylor Barrella, Xi Dong, Sean A. Hartnoll and Victoria L. Martin

\vspace{1cm}

{\small
{\it Department of Physics, Stanford University, \\
Stanford, CA 94305-4060, USA }}

\vspace{1.6cm}

\end{center}

\begin{abstract}

The R\'enyi entropies and entanglement entropy of 1+1 CFTs with gravity duals can be computed by explicit construction of the bulk spacetimes dual to branched covers of the boundary geometry. At the classical level in the bulk this has recently been shown to reproduce the conjectured Ryu-Takayanagi formula for the holographic entanglement entropy. We study the one-loop bulk corrections to this formula. The functional determinants in the bulk geometries are given by a sum over certain words of generators of the Schottky group of the branched cover. For the case of two disjoint intervals on a line we obtain analytic answers for the one-loop entanglement entropy in an expansion in small cross-ratio. These reproduce and go beyond anticipated universal terms that are not visible classically in the bulk. We also consider the case of a single interval on a circle at finite temperature. At high temperatures we show that the one-loop contributions introduce expected finite size corrections to the entanglement entropy that are not present classically. At low temperatures, the one-loop corrections capture the mixed nature of the density matrix, also not visible classically below the Hawking-Page temperature.

\end{abstract}

\pagebreak
\setcounter{page}{1}

\tableofcontents

\pagebreak

\section{Holographic entanglement entropy}

The entropic nature of black holes \cite{Bekenstein:1973ur, Hawking:1974sw} has hinted for several decades now that a fundamental theory of gravity may involve information processing in an essential way. The Bekenstein-Hawking connection between spacetime and information was substantially generalized seven years ago by the provocatively simple proposal of Ryu and Takayanagi for the gravitational description of the entanglement entropy in field theories with holographic duals \cite{Ryu:2006bv}. In the simplest setting, their proposal conjectured that the entanglement entropy of a spatial region in the field theory is given by the area of a minimal surface in the dual bulk geometry that extends to the conformal boundary of the bulk spacetime and whose boundary is that of the spatial region of interest. This statement is simultaneously a concrete step towards the reformulation of spacetime as entanglement and also an efficient tool for the computation of entanglement entropy in certain strongly interacting systems.

The Ryu-Takayanagi proposal has very recently been proven in detail for certain 1+1 dimensional conformal field theories with large central charge \cite{Hartman:2013mia, Faulkner:2013yia}, while strong arguments for its validity in higher dimensions have also been presented \cite{Lewkowycz:2013nqa}. These papers built on earlier works, some of which we shall mention below. In the case of 1+1 dimensional conformal field theories (CFTs) with gravity duals, a completely explicit construction of the bulk spacetimes needed to compute the entanglement R\'enyi entropies was achieved \cite{Faulkner:2013yia}. In this paper we shall take these results as a starting point to compute bulk quantum corrections to the entanglement R\'enyi entropies. Analytic continuation of the R\'enyi entropies allows us to obtain the bulk one-loop corrections to the entanglement entropy in these theories. The one-loop correction is perturbatively exact in pure three dimensional gravity \cite{Maloney:2007ud}. In this way we start an exploration of holographic entanglement beyond the classical gravity regime of validity of the Ryu-Takayanagi formula.

In this paper we will focus on three properties of entanglement in 1+1 dimensional CFTs that are not visible to leading order in the holographic large central charge expansion. By obtaining one-loop corrections to the Ryu-Takayanagi formula,
we shall demonstrate explicitly that these properties are instead exhibited at the one-loop level in the bulk. This is achieved by building on formulae for functional determinants in quotients of $AdS_3$ that were derived by \cite{Giombi:2008vd}. The three properties we study are
\begin{enumerate}

\item Consider two disjoint intervals in the CFT on a line with the distance between the intervals much larger than the length of the intervals. The Ryu-Takayanagi minimal surface becomes two soap bubbles ending on each interval separately. There is no mutual information. It is known, however, that there are universal terms that must appear in the entanglement entropy that depend on the distance between the two intervals \cite{Headrick:2010zt, Calabrese:2010he}. We (re)derive exactly the universal terms, as well as additional terms, from bulk one-loop contributions to the entanglement entropy.

\item Consider a single interval in a CFT on a circle and at a finite temperature above the Hawking-Page temperature. The Ryu-Takayanagi holographic entanglement entropy has previously been computed from the corresponding geodesics in the BTZ black hole background \cite{Ryu:2006ef} and the result found to agree with a universal formula for the entanglement entropy of an interval in a finite temperature CFT on a {\it line} \cite{Holzhey:1994we}. The finite size corrections to the entanglement, that induce deviations from the universal formula on a line, are shown to appear in loop corrections in the bulk.

\item Consider a single interval in a CFT on a circle and at a finite temperature below the Hawking-Page temperature. Because the system is in a mixed state, the entanglement entropy of the interval and its complement will generally not be equal. However, below the Hawking-Page temperature, the bulk geometry is thermal $AdS_3$ and no remnant of finite temperature effects are seen in the Ryu-Takayanagi entanglement (this point was emphasized in \cite{Herzog:2012bw}). We show that bulk one-loop corrections to the entanglement entropy do generate the expected asymmetry between the interval and its complement.

\end{enumerate}

Our discussion of the CFT on a circle and finite temperature will involve a generalization of the uniformization map used in \cite{Hartman:2013mia, Faulkner:2013yia} to the case of branched covers of a torus. Indeed, we will start by recalling the connection between the holographic entanglement entropy of CFTs and the Schottky uniformization of Riemann surfaces.

As well as the one-loop entanglement entropy, we also obtain new analytic results -- in the limits discussed above -- for the R\'enyi entropies at a classical level in the bulk. We find that certain information in the classical R\'enyi entropies (mutual information of well separated intervals, finite size effects at high temperatures) drops out in the $n \to 1$ limit in which the classical bulk contribution to the entanglement entropy is obtained. This simplification may help to decode the connection between entanglement entropy and spacetime geometry.

Looking towards the future, we hope that the various analytic results obtained in this paper can serve as useful data points for a possible reformulation of the full quantum bulk theory in terms of entanglement entropies.

\section{From entanglement entropy to Schottky uniformization}

In this section we summarize results relating the entanglement entropy in 1+1 dimensional CFTs to the partition function of the CFT on higher genus Riemann surfaces. We further review how Schottky uniformization of the Riemann surface relates these partition functions to gravity on specific quotients of $AdS_3$.

The entanglement between degrees of freedom inside and outside of a spatial region $A$ is characterized by the reduced density matrix $\rho$ associated to this region. The entanglement R\'enyi entropies in particular are defined as
\be
S_n = - \frac{1}{n-1} \log \, \tr \, \rho^n \,.
\ee
The entanglement entropy itself is then obtained by analytic continuation of the R\'enyi entropies to non-integer $n$ and taking the limit
\be
S = \lim_{n \to 1} S_n = - \tr \, \rho \log \rho \,.
\ee

In 1+1 dimensions the spatial region $A$ is given by a set of disjoint intervals. The $n$th trace of the density matrix is then equal to the partition function of the 1+1 dimensional CFT evaluated on an $n$-sheeted cover of the original spacetime, with branch cuts running between the pairs of points delimiting the disjoint intervals (see e.g. \cite{Nishioka:2009un}). Explicitly,
\be\label{eq:sz}
S_n = - \frac{1}{n-1} \log \,\frac{Z_n}{Z_1^n} \,.
\ee
Here $Z_1$ is the partition function of the CFT on the initial spacetime in which the theory was defined and $Z_n$ is the partition function on the $n$-sheeted cover. The cover defines a higher genus Riemann surface $\Sigma$. Discussions of these surfaces in certain cases, including the resolution of the conical singularities at the branch points, can be found in e.g. \cite{Lunin:2000yv, Headrick:2010zt, Hung:2011nu, Faulkner:2013yia}. The immediate objective is to evaluate the partition functions $Z_n$.

In the $AdS_3$/CFT$_2$ correspondence, the partition function of the CFT on a Riemann surface $\Sigma$ is equal to the partition function of the dual gravitational theory on a quotient $AdS_3/\Gamma$ with conformal boundary $\Sigma$. The quotient is by a discrete subgroup $\Gamma \subset PSL(2,\C)$, the isometry group of $AdS_3$. The conformal boundary inherits a quotient action. In particular, if the metric on $AdS_3$ is written
\be\label{eq:ads3}
ds^2 = \frac{d\xi^2 + d w d \bar w}{\xi^2} \,,
\ee
then near the conformal boundary $\xi \to 0$, the elements of $PSL(2,\C)$ act as M\"obius transformations on the conformal boundary
\be\label{eq:mobius}
w \mapsto L(w) \equiv \frac{a w + b}{c w + d} \,, \qquad \xi \mapsto |L'(w)| \xi \,, \qquad ad - bc = 1\,.
\ee
The symmetry action on the full bulk spacetime can be found in e.g. \cite{Faulkner:2013yia}. To evaluate the bulk partition function we therefore need to find a subgroup $\Gamma$ such that our $n$-sheeted Riemann surface $\Sigma = \C/\Gamma$, with $\Gamma$ now acting via the M\"obius transformations (\ref{eq:mobius}). Strictly, we need to remove from $\C$ the fixed points of $\Gamma$ before taking the quotient. Finding this representation, that is to say, realizing $\Sigma$ as a quotient $\Sigma = \C/\Gamma$, with certain restrictions on $\G$, amounts to a Schottky uniformization of $\Sigma$.

It is a theorem (see e.g. \cite{schottky} for discussion and references) that every compact Riemann surface can be obtained as the quotient $\Sigma = \C/\Gamma$ with $\Gamma$ a Schottky group. The Schottky group of a genus $g$ Riemann surface is a subgroup of $PSL(2,\C)$ that is freely generated by $g$ loxodromic elements of $PSL(2,\C)$. The connection between the genus and the group goes as follows. Write the $g$ generators of the Schottky group as $\{L_i\}_{i=1}^g$. M\"obius transformations map circles to circles and in particular, for these loxodromic transformations, $2g$ disjoint circles $\{C_i, C_i'\}_{i=1}^g$ can be chosen such that $L_i(C_i) = C_i'$. Under the quotient $\Sigma = \C/\Gamma$ these circles in $\C$ map to $g$ nontrivial elements of the fundamental group $\pi_1(\Sigma)$. Specifically, the circles generate a maximal freely generated subgroup of the fundamental group. The remaining $g$ generators of the fundamental group are then obtained by paths that connect the pairs of circles. For more details of this construction see e.g. \cite{schottky, Krasnov:2000zq,Faulkner:2013yia} and references therein.

There will typically be more than one quotient of $AdS_3$ that has a given Riemann surface $\Sigma$ as conformal boundary. In particular it is not proven that the dominant bulk geometry realizes a Schottky uniformization of the Riemann surface. Quotients of $AdS_3$ by non-Schottky groups give non-handlebody bulk geometries. See e.g. \cite{Yin:2007gv, Yin:2007at}. Following the results in \cite{Faulkner:2013yia}, it seems to be the case that the dominant contributions are in fact given by quotients by Schottky groups. We will assume this in the following.

The strategy to obtain the $n$th R\'enyi entropy is therefore as follows: (i) Find a Schottky uniformization of the corresponding branched cover $\Sigma = \C/\Gamma$ and then (ii) compute the partition function of the dual gravitational theory on the associated quotient $AdS_3/\Gamma$.

\newpage

\section{Schottky uniformization of branched covers of $\C$ and $T^2$}
\label{sec:unif}

Throughout this paper we focus on two illustrative cases. The entanglement entropy of two intervals on a line at zero temperature and the entanglement entropy of one interval on a circle at finite temperature.

\subsection{Two intervals on a plane}
\label{sec:twointervals}

Two disjoint intervals on a plane constitute the simplest setting in which the entanglement entropy of a 1+1 CFT is not determined entirely by the central charge \cite{Calabrese:2009ez, Calabrese:2010he}. At the classical bulk level, this case was considered in detail holographically in \cite{Faulkner:2013yia}. Below we obtain one-loop corrections in the bulk that capture the mutual information between the two intervals that is not visible classically.

Let $z$ be the complex coordinate on the plane. Let the two intervals be bounded by the four real numbers $\{z_i\}_{i=1}^4$.
To obtain the $n$th R\'enyi entropy we must compute the partition function of the $n$-sheeted complex plane with branch points at the four $z_i$. The first step in the Schottky uniformization is to define coordinates $w$ that are single-valued on the $n$-sheeted cover. This is achieved by considering the differential equation
\be\label{eq:diffeq}
\psi''(z) + \frac{1}{2} \sum_{i=1}^4 \left(\frac{\Delta}{(z-z_i)^2} + \frac{\g_i}{z-z_i} \right) \psi(z) = 0 \,,
\ee
with
\be\label{eq:Delta}
\Delta = \frac{1}{2} \left( 1 - \frac{1}{n^2} \right) \,.
\ee
This value of $\Delta$ is chosen, as we see immediately below, to fix the behavior of solutions to the equation near the branch points in a way that will allow us to construct a singe-valued coordinate on the $n$-sheeted cover.
The four $\g_i$ are called accessory parameters and are to be fixed shortly. Take two independent solutions $\{ \psi_1, \psi_2 \}$ to (\ref{eq:diffeq}) and write
\be\label{eq:wdef}
w = \frac{\psi_1(z)}{\psi_2(z)} \,.
\ee
Near to each of the branch points $z_i$, the solutions to (\ref{eq:diffeq}) behave as $(z-z_i)^{(1 \pm 1/n)/2}$. Therefore $w(z)$ can be written as a power series expansion in $(z-z_i)^{1/n}$. It follows that $w$ is single-valued on the $n$-sheeted plane.

If we follow the pair of independent solutions $\{ \psi_1, \psi_2 \}$ around a closed loop $C$ enclosing one or more branch points, the solutions will generically experience some monodromy
\be\label{eq:monod}
\left(
\begin{array}{c}
\psi_1 \\
\psi_2
\end{array}
\right) \mapsto
M(C) \left(
\begin{array}{c}
\psi_1 \\
\psi_2
\end{array}
\right) \,, \qquad M(C) =
\left(
\begin{array}{cc}
a & b \\
c & d
\end{array}
\right) \in PSL(2,\C) \,.
\ee
From the definition (\ref{eq:wdef}) of $w$, it is immediate that the monodromies (\ref{eq:monod}) induce $PSL(2,\C)$ identifications on the $w$ coordinate
\be
w \sim \frac{a \, w + b}{c \, w + d} \,.
\ee
Therefore, starting with the complex $w$ plane, we can find a Schottky uniformization of the $n$-sheeted cover. We must be able to fix the accessory parameters so that the monodromy identifications around $g = n - 1$ non-intersecting cycles are nontrivial, while the monodromy transformations around their dual cycles all lead to trivial identifications.

In stating the genus we have implicitly compactified the complex plane at infinity. The compactification amounts to requiring a trivial monodromy at infinity. Expanding the differential equation (\ref{eq:diffeq}) at large $z$, this is seen to impose
\be\label{eq:conditions}
\sum_{i=1}^4 \g_i = 0 \,, \qquad \sum_{i=1}^4 \g_i z_i = - 4 \Delta \,, \qquad \sum_{i=1}^4 \g_i z_i^2 = - 2 \Delta \sum_{i=1}^4 z_i \,.
\ee
This fixes three of the four accessory parameters. A trivial monodromy at infinity leaves only two homotopically distinct cycles on each sheet. These can be taken, for instance, to be a cycle enclosing $[z_1,z_2]$ and another enclosing $[z_2,z_3]$. We must fix the remaining accessory parameter by imposing that the monodromy around one of these two cycles be trivial. In later sections of this paper we shall do this both numerically and, in certain regimes, analytically. Which cycle to trivialize is a choice of Schottky uniformization. We will see below that this is determined dynamically. The remaining nontrivial monodromy on each sheet then gives the $g = n-1$ generators $L_i$ of the Schottky group. There is one generator per sheet, except that the monodromies on the final sheet are not independent of the previous ones. To be more precise, let us denote the monodromy obtained by encircling one of the branch points delimiting the nontrivial cycle by $M_1$ and the monodromy obtained by encircling the other branch point by $M_2$. These monodromies do not correspond to closed paths on the Riemann surface, as encircling a single branch point moves us between sheets. We then see that the $n-1$ generators of the Schottky group are
\be\label{eq:othergen}
L_1 = M_2 M_1 \,, \qquad L_i = M_2^{i-1} L_1 M_2^{-(i-1)} = M_2^i M_1 M_2^{-(i-1)} \,, \qquad i = 2, \ldots, n-1.
\ee
It follows that $L_n L_{n-1} \ldots L_1 = 1$ and therefore $L_n$ is not independent. This last statement requires use of the fact that $M_1^n = M_2^n = -1$, as can be verified directly by expanding solutions to the equation (\ref{eq:diffeq}) about the branch points. We will see this explicitly below. The generators (\ref{eq:othergen}) are illustrated in figure \ref{fig:schot} below, for two choices of Schottky group.
\begin{figure}[h]
\centering
\includegraphics[width=0.95\textwidth]{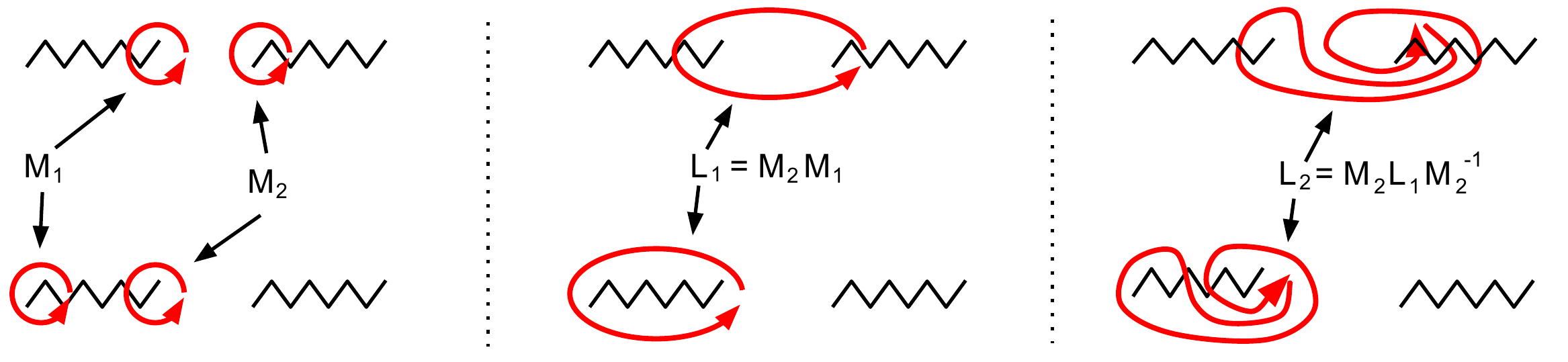}   
\caption{\label{fig:schot} Some of the cycles in (\ref{eq:othergen}) generating the Schottky group for two choices of uniformization. In the top line the monodromy around the $[z_1,z_2]$ cycle has been trivialized and so the generators use the remaining $[z_2,z_3]$ cycle. In the second line the $[z_2,z_3]$ cycle has been trivialized and the generators use the remaining $[z_1,z_2]$ cycle.}
\end{figure}

As was stressed in \cite{Faulkner:2013yia}, in writing the equation (\ref{eq:diffeq}) we have assumed that the Schottky uniformization leading to the dominant bulk geometry respects the $\Z_n$ `replica' symmetry. Technically this means that we have not included otherwise allowed terms in (\ref{eq:diffeq}) that violate this symmetry. The general theory of accessory parameters is summarized in \cite{schottky}.

Finally, note that we start with the flat metric on the original complex $z$ plane. This flat metric is inherited by the branched cover. The change of variables $w(z)$ in (\ref{eq:wdef}) will induce a non-flat metric on the complex $w$ plane. Our discussion relating Schottky uniformization to an $AdS_3$ bulk around equation (\ref{eq:ads3}), in contrast, assumed a flat metric on the $w$ plane. This is not a problem: the metric on the $w$ plane may be made flat via a Weyl transformation. The effect of this transformation on the partition function is entirely captured by the conformal anomaly and so is easily computable.

\subsection{One interval on a torus}

A single interval on a circle at finite temperature is of interest to us because it provides another simple setting in which a qualitatively important effect is not visible at leading order in the bulk. At finite temperature the density matrix of the whole system is in a mixed state. It follows that the entanglement entropy of a region and its complement will not be equal. Above the Hawking-Page temperature, this effect is visible in classical gravity because the minimal surfaces describing a region and its complement wrap different sides of the bulk black hole \cite{Ryu:2006bv}. However, below the transition temperature there is no black hole in the spacetime and the minimal surfaces are identical.

Above the Hawking-Page temperature there is also a qualitative effect that is missing at leading order in the bulk. The entanglement entropy is found to be independent of the length of the spatial circle. In terms of entanglement entropy at least, it is as though the Hawking-Page transition were between zero temperature and infinite temperature phases.

Finite temperature corresponds to periodically identifying the Euclidean time circle. The spatial dimension is also taken to be a circle and therefore the partition functions of interest are on branched covers of a torus. Let $z$ be the coordinate on the torus, with periodicities
\be\label{eq:period}
z \sim z + R \, \Z + \frac{i \, \Z}{T} \,.
\ee
We think of $R$ as the length of the spatial circle and $T$ as the temperature. The differential equation we are looking for in order to uniformize the $n$-sheeted cover must now have the following properties. Firstly, it must have the same behavior near the branch points $\{z_i\}_{i=1}^2$ as in the previous equation (\ref{eq:diffeq}), so that the coordinate $w(z)$ in (\ref{eq:wdef}) is again single-valued on the branched cover. Secondly the terms in the equation should be elliptic functions. Retaining the $\Z_n$ replica symmetry, these conditions fix the differential equation up to a single accessory parameter $\g = \g_1 = - \g_2$ and an additional constant $\d$ to be
\be\label{eq:diffeqtorus}
\psi''(z) + \frac{1}{2} \sum_{i=1}^2 \Big(\Delta \, \wp(z-z_i) + \g (-1)^{i+1} \zeta(z-z_i) + \d \Big) \psi(z) = 0 \,,
\ee
with $\Delta$ given as before in (\ref{eq:Delta}). Here $\wp$ is the Weierstrass elliptic function and $\zeta$ is the Weierstrass zeta function.  The definitions and details of these functions are summarized in Appendix \ref{sec:weierstrass}.  Note that while $\z(z)$ is not an elliptic function, the combination $\zeta(z-z_1) - \zeta(z-z_2)$ that appears is indeed doubly periodic. The fact that the simple poles of elliptic functions necessarily come in pairs with opposite residues is what constrains the differential equation to have only a single accessory parameter. Here and below we are suppressing the periods of the Weierstrass functions, which are clearly those compatible with the torus identifications (\ref{eq:period}).
The need for the constant term $\d$ in the equation can be seen by considering the torus in the absence of a branch cut, with $\Delta = \g = 0$. In that case, in order to trivialize the monodromy around either the spatial or the time circle -- and thereby obtain Schottky uniformizations of the torus -- we must have $\d = \pi^2/R^2$ or $\d = -(\pi T)^2$, respectively. Note that the monodromy matrix in these cases is minus the identity, which is equal to the identity in $PSL(2,\Z)$.

Once $\Delta \neq 0$, we Schottky uniformize the $n$-sheeted cover of the torus by choosing $\g$ and $\d$ so that the monodromy is trivial around both the spatial or temporal circle of the torus and also the cycle that encloses the branch cut between $z_1$ and $z_2$. The remaining monodromy around the time or space circle, together with a construction analogous to (\ref{eq:othergen}) for the monodromy on the higher sheets, will then determine the action of the Schottky group on the $w$ plane. We will perform this construction explicitly in later sections.

\section{Entanglement entropy in classical gravity}
\label{sec:classical}

To leading order at large central charge the partition function is holographically given by the on-shell Euclidean action of the dual geometry
\be\label{eq:action}
- \log Z = S_E = - \frac{1}{2\k^2} \int d^3x \sqrt{g} \left(R + \frac{2}{L^2} \right)
- \frac{1}{\k^2} \int d^2x \sqrt{\gamma} \left(K - \frac{1}{L} \right) \,. 
\ee
The bulk term is the Einstein-Hilbert action with a negative cosmological constant that sets the $AdS_3$ radius to $L$. The boundary terms are the usual Gibbons-Hawking and boundary counterterm actions. This action must be evaluated on $AdS_3/\Gamma$, with $\Gamma$ obtained via the Schottky uniformization outlined in the previous section.

The on-shell action was evaluated in \cite{Faulkner:2013yia}, borrowing heavily from \cite{schottky} and \cite{Krasnov:2000zq}, where it was expressed in terms of the accessory parameters characterizing the Schottky uniformization:
\be\label{eq:dsdz}
\frac{\pa S_E}{\pa z_i} = - \frac{c \, n}{6} \gamma_i \,.
\ee
Here the central charge, see e.g. \cite{Balasubramanian:1999re} and references therein,
\be
c = \frac{12 \pi L}{\k^2} \,.
\ee
The derivative of the entanglement entropy is therefore given by
\be\label{eq:sofz}
\frac{\pa S}{\pa z_i} = - \lim_{n \to 1} \frac{c \, n}{6 (n-1)} \gamma_i \,.
\ee
Here $\gamma_i(n)$ is easily analytically continued to $n$ non-integer because it is defined by requiring the absence of certain monodromies of a differential equation, as we described in section \ref{sec:unif} above. This differential equation is well-defined for all $n$.

The computation of the action is set up as follows. Given the metric on $AdS_3$ in (\ref{eq:ads3}), introduce a cutoff surface at
\be\label{eq:cutoff}
\xi_c = \frac{\epsilon}{L} e^{-\phi(w,\bar w)/2} \,.
\ee
Here $\epsilon \ll 1$ regulates the infinite volume divergence of the on-shell action. The field $\phi(w,\bar w)$ will be specified shortly. In addition to this cutoff, the $AdS_3$ geometry is quotiented by the Schottky group $\G$. As reviewed in e.g. \cite{Krasnov:2000zq, Faulkner:2013yia}, the bulk action of $PSL(2,\C)$ maps hemispheres to hemispheres. At the conformal boundary these hemispheres end on circles. As we recalled in the main text, the Schottky uniformization is characterized by $g$ pairs of circles $\{C_i,C'_i\}_{i=1}^g$ that are identified under the action of $\G$. These circles extend to hemispheres in the bulk that are identified. Without loss of generality we can take one of these circles $C_1$ on the boundary to enclose the remaining $2g-1$ circles. The bulk action is proportional to the volume of $AdS_3/\Gamma$ with the cutoff (\ref{eq:cutoff}), which can now be written as
\be\label{eq:UVIR}
\int d^3x \sqrt{g} = L^3 \int_D d^2x \int_{\xi_\text{IR}(w,\bar w)}^{\xi_\text{UV}(w,\bar w)} \frac{d \xi}{\xi^3} \,.
\ee
Here $D$ is the disc inside the largest circle $C_1$ on the boundary. The IR cutoff $\xi_\text{IR}(w,\bar w)$ is given by the hemisphere ending on $C_1$ at the boundary. The UV cutoff $\xi_\text{UV}(w,\bar w)$ is given by $\xi_c$ in the fundamental region of the Schottky group, that is, outside of any of the remaining circles, whereas inside any of the remaining circles it is given by the hemisphere ending on that circle.

To leading order as $\epsilon \to 0$, the induced boundary metric is $ds^2 = e^{\phi} dw d\bar w$. The objective is to compute the partition function on the branched covers with metric $ds^2 = dz d\bar z$. We therefore set
\be\label{eq:phidef}
e^{-\phi} = \left| \frac{\pa w}{\pa z} \right|^2 \,.
\ee
Recall that $w(z)$ is given by the uniformization map (\ref{eq:wdef}). Therefore (\ref{eq:phidef}) defines the Liouville field $\phi$ in terms of $w,\bar w$. The bulk action with cutoff specified by this function $\phi$ computes the semiclassical contribution to the $n$th R\'enyi entropy that we are after.

On-shell, the gravitational action (\ref{eq:action}) becomes a boundary term. From the discussion around equation (\ref{eq:UVIR}), we see that there will be two types of boundary terms. There are those evaluated on the cutoff $\xi = \xi_c$ and those evaluated on the hemispheres associated with the bulk quotient action. The computation in \cite{Faulkner:2013yia}, to which we refer the reader, evaluated all of these terms in the action and obtained the result (\ref{eq:dsdz}).

The expression (\ref{eq:dsdz}) implies that the entanglement and R\'enyi entropies are given by the accessory parameters, which are in turn found in the process of obtaining the Schottky uniformization of the branched cover. The Ryu-Takayanagi formula for the entanglement entropy was derived from these expressions in the case of two intervals on a line in \cite{Faulkner:2013yia}. In the remainder of this section we will obtain the entanglement entropy for the case of one interval on a torus in the classical bulk limit.

\subsection{One interval on a torus: classical result}
\label{sec:tclassical}

The objective is to obtain the accessory parameter $\gamma$ (as well as the constant $\delta$) in the torus differential equation (\ref{eq:diffeqtorus}). Given the accessory parameter, we will obtain the classical gravity contribution to the entanglement entropy from (\ref{eq:sofz}). While this formula was only proven in \cite{Faulkner:2013yia} for intervals on a line, we expect that the derivation will go through on a finite temperature circle also. Indeed, in support of this last statement, we will reproduce the Ryu-Takayanagi result for an interval on a torus.

Expanding in $\vep \equiv n-1$ we write
\be\label{eq:nminus1}
\psi(z) = \psi_{(0)}(z) + \vep \psi_{(1)}(z) \,, \qquad \Delta = \vep \,, \qquad \g = \vep \g_{(1)} \,, \qquad \delta = - (\pi T)^2 + \vep \d_{(1)} \,.
\ee
The choice of $\delta$ here means that we are trivializing the monodromy around the time circle, rather than the spatial circle. This corresponds to bulk geometries in which the time circle is contractible, which holds for temperatures above the Hawking-Page transition. The zeroth order solution is easily obtained
\be\label{eq:zerothorder}
\psi_{(0)}(z) = A e^{z \pi T} + B e^{- z \pi T} \,.
\ee
Here $A$ and $B$ are constants of integration. The choice of $\delta = - (\pi T)^2$ at zeroth order has ensured that these solutions have no monodromy (in $PSL(2,\Z)$) around the time circle.

Without loss of generality, we will take $z_1 = y$ and $z_2 = R - y$. The branch cut from $z_1$ to $z_2$ is going `around' the torus. This is a convenient choice because we can now take the cycle around the time circle to be at $\text{Re} \, z = 0$, which does not cross the branch cut.

The first order solution is also straightforward to obtain. The homogeneous part can be absorbed into $\psi_{(0)}$, while the inhomogeneous part is given by
\be\label{eq:firstorder}
\psi_{(1)}(z) = \frac{e^{-z \pi T}}{2 \pi T} \int_0^{z} e^{x \pi T} m(x) \psi_{(0)}(x) dx -  \frac{e^{z \pi T}}{2 \pi T} \int_0^{z} e^{- x \pi T} m(x) \psi_{(0)}(x) dx \,,
\ee
where
\be
m(z) =  \frac{1}{2} \sum_{i=1}^2 \Big(\wp(z-z_i) + \g_{(1)} (-1)^{i+1} \zeta(z-z_i) + \delta_{(1)} \Big) \,.
\ee
We need to impose that the first order solution does not introduce any monodromy around the time circle. This will be the case if, for all zeroth order solutions $\psi_{(0)}$,
\be
\psi_{(1)}(0) = \psi_{(1)}'(0) = \psi_{(1)}(i/T) = \psi_{(1)}'(i/T) = 0\,.
\ee
From (\ref{eq:zerothorder}) and (\ref{eq:firstorder}), these conditions are seen to be equivalent to the requirements
\be\label{eq:integrals}
\int_0^{1} m(i s/T) ds = 0 \,, \qquad \int_0^{1} e^{\pm2 \pi i s} m(i s/T) ds = 0 \,.
\ee
In fact $m$ is an even function (with the $z_i$ as chosen above) and so the sign in the exponent of the second relation is not important.

The first integral in (\ref{eq:integrals}) is simple to perform and gives
\be
\delta_{(1)} = \frac{T}{2 i} \left(\zeta(i/T-y) + \zeta(i/T + y) + \g_{(1)} \log \frac{\sigma(i/T + y-R) \sigma(-y)}{\sigma(y-R) \sigma(i/T - y)} \right) \,.
\ee
Here $\sigma$ is the Weierstrass sigma function. The accessory parameter $\g_{(1)}$ is now obtained from the second integral in (\ref{eq:integrals}). First we note that, integrating the $\zeta$ function terms in $m(z)$ by parts, we obtain
\be
\sum_i \left(1 + \frac{\g_{(1)} (-1)^{i+1}}{2 \pi T} \right) \int_0^1 e^{2 \pi i s} \wp(i s/T - z_i) ds = 0 \,.
\ee
This last integral can be performed by contour integration
\be
\int_0^1 e^{2 \pi i s} \wp(i s/T - z_i) ds = \frac{(2 \pi T)^2 e^{2 \pi T z_i}}{e^{2 \pi T R}-1} \,.
\ee
We thereby obtain
\be
\g_{(1)} =  2 \pi T \coth \pi T (z_2 - z_1) \,.
\ee
From (\ref{eq:sofz}), the entanglement entropy is therefore
\be\label{eq:storus}
S = \frac{c}{6} \log \sinh^2 \pi T (z_2 - z_1) + \text{const} \,.
\ee

We can compare the result (\ref{eq:storus}) to the Ryu-Takayanagi result. This was obtained in \cite{Ryu:2006ef} from the lengths of geodesics in the BTZ background. We can write the BTZ metric as
\be
ds^2 = L^2 \left(-(r^2 - r_+^2) dt^2 + \frac{dr^2}{r^2 - r_+^2} + r^2 d\phi^2 \right) \,,
\ee
where $\phi \sim \phi + R$ and $t \sim t + T^{-1} = t + 2\pi/r_+$. The proper length of the fixed time geodesic 
connecting two points with separation $\Delta\phi$ at the conformal boundary is easily computed. This gives the holographic entanglement entropy
\be\label{eq:hightclass}
S = \frac{\text{length}}{\k^2/2\pi} = \frac{c}{6} \log \left( \frac{4 r_c^2}{r_+^2} \sinh^2 \frac{r_+ \Delta \phi}{2} \right) \,.
\ee
Here $r_c$ is the UV cutoff. Up to non-universal cutoff terms, we obtain an exact agreement with the result (\ref{eq:storus}) from Schottky uniformization.

As we noted in the introduction, the result (\ref{eq:storus}) can be derived for a general CFT to describe the entanglement entropy of an interval on a line at finite temperature \cite{Holzhey:1994we}. On a circle, we would expect (\ref{eq:storus}) to arise only in the high temperature limit. We will obtain finite size corrections to the entanglement entropy below at one loop in the bulk. The absence of finite size corrections to leading order in large central charge seems to be a specific feature of CFTs with gravity duals. This is likely related to the fact that the asymptotic Cardy formula reproduces the exact BTZ black hole entropy at all temperatures \cite{Strominger:1997eq}. Interestingly, in this regard, we will see in section \ref{sec:tloop} below that the R\'enyi entropies with $n>1$ do contain finite size corrections already at the classical level. It is only the entanglement entropy proper that misses these features to leading order.

Below the Hawking-Page transition we need to trivialize the monodromy around the spatial circle. Thus we expand the quantities in $n-1$ as in (\ref{eq:nminus1}), except that now $\d = (\pi/R)^2 + (n-1)\delta_{(1)}$. The computation goes through as before, with some factors of $i$ in different places. The answer for the entanglement entropy is found to be
\be\label{eq:lowT}
S = \frac{c}{6} \log \sin^2 \frac{\pi \Delta \phi}{R} + \text{const} \,.
\ee
This is the universal answer for the entanglement entropy of an interval in a CFT on a circle at zero temperature \cite{Holzhey:1994we} and also agrees with the Ryu-Takayanagi formula \cite{Ryu:2006ef}. Just as the result (\ref{eq:storus}) above the Hawking-Page transition did not know about the finite size of the spatial circle, the result (\ref{eq:lowT}) does not know about the temperature. In particular, it is invariant under $\Delta \phi \to R - \Delta \phi$. Therefore the classical bulk result for the entanglement entropy does not capture the expected difference in the entaglement between a region and its complement at nonzero temperature.

\section{One-loop correction to the entanglement entropies}
\label{sec:oneloop}

In the semiclassical expansion of the partition function at large central charge, the leading correction to the on-shell action of the previous section is given by the functional determinant of the operator describing quadratic fluctuations of all the bulk fields. This one-loop correction is perturbatively exact in pure three dimensional gravity \cite{Maloney:2007ud}. Fortunately, an elegant expression for functional determinants on quotients of $AdS_3$ by a Schottky group $\G$ has been obtained in \cite{Giombi:2008vd}, following a conjecture in \cite{Yin:2007gv}. The answer for metric fluctuations is found to be
\be\label{eq:zoneloop}
\left. \log Z \right|_\text{one-loop} = - \sum_{\g \in {\mathcal P}} \sum_{m=2}^\infty \log |1 - q_\g^m| \,.
\ee
Here ${\mathcal P}$ is a set of representatives of the primitive conjugacy classes of $\G$. Recall that an element $\g \in \G$ is primitive if it cannot be written as $\b^n$ for any element $\b \in \G$ and $n > 1$. In (\ref{eq:zoneloop}), $q_\gamma$ is defined by writing the two eigenvalues of $\g \in \G \subset PSL(2,\C)$ as $q_\g^{\pm 1/2}$ with $|q_\g| < 1$. A similar expression also exists for other bulk fields. For instance, for scalar fields dual to operators with scaling dimension $h$, the contribution is \cite{Giombi:2008vd}
\be\label{eq:scalarloop}
\left. \log Z \right|_\text{one-loop} = - \frac{1}{2} \sum_{\g \in {\mathcal P}} \sum_{\ell,\ell'=0}^\infty \log \left(1 - q_\g^{\ell + h/2} \bar q_\g^{\ell' + h/2} \right) \,.
\ee
In addition to the expression given in these sums, the one-loop contribution renormalizes the bulk cosmological constant.

We will evaluate these determinants and obtain the one-loop contribution to the entanglement entropy using the following steps:
\begin{enumerate}
\item Find the Schottky group $\G$ corresponding to the $n$-sheeted covers. This involves solving the monodromy problem described in the previous sections and then obtaining the generators $L_i$ of the group as in (\ref{eq:othergen}).

\item Generate ${\mathcal P}$ for the Schottky group $\G$, by forming non-repeated words from the $L_i$ and their inverses, up to conjugation in $\G$. There are infinitely many such primitive conjugacy classes for genus $g>1$ (corresponding to $n > 2$ for two intervals on the plane and $n > 1$ for one interval on the torus).

\item Compute the eigenvalues of these words and thereby compute the infinite sums appearing in e.g. (\ref{eq:zoneloop}).

\item To obtain the entanglement entropy from the entanglement R\'enyi entropies, analytically continue the one-loop contribution to $S_n$ to $n \to 1$.

\end{enumerate}
In order to perform the final step of analytic continuation, analytic results for the R\'enyi entropies as a function of $n$ are necessary. The analytic continuation here is trickier than for the classical contribution to the entropy discussed in section \ref{sec:classical}. This is because while the accessory parameters can be directly computed at non-integer $n$, via the uniformizing differential equations such as (\ref{eq:diffeq}), the sum over elements of ${\mathcal P}$ is only defined at integer $n$. One must therefore perform the sum explicitly for each integer $n$ and then find a way to analytically continue the result. We have achieved this in certain limits. Outside of these limits, we can compute the R\'enyi entropies numerically.

\section{Two intervals on a line: small cross-ratio expansion}
\label{sec:univ}

We are able to compute the one-loop contribution to the entanglement entropy analytically in an expansion in the cross-ratio
\be\label{eq:cross}
x \equiv \frac{(z_3-z_2)(z_4-z_1)}{(z_3-z_1)(z_4-z_2)} \,.
\ee
The one-loop entropies will only depend on this combination of the coordinates of the endpoints of the intervals. This property is inherited, for the one-loop contribution, from the mutual information as we discuss in section \ref{sec:twoloop} below.
Using conformal invariance, without loss of generality we can place the locations of the intervals at $(z_1,z_2,z_3,z_4) = (-1,-y,y,1)$ with $0 < y < 1$. This will be useful for intermediate computations.  Let us consider the entanglement associated to the two intervals $[-y,y]$ and $[1,-1]$, where by the interval $[1,-1]$ we mean $[1,\infty)\cup (-\infty,-1]$.  The cross-ratio $x$ is related to $y$ as
\be
x=\frac{4y}{(y+1)^2} \, .
\ee
Small $x$ therefore corresponds to small $y$.

The leading order result in an expansion in small $x$ is known in general. The R\'enyi entropies are given by \cite{Calabrese:2010he}
\beq\label{s1nx}
S_n = - \mathcal N \frac{n}{2(n-1)} \left(\frac{x}{4 \, n^2}\right)^{2 h} \sum_{k=1}^{n-1} \frac{1}{\[\sin(\pi k/n)\]^{4 h}}+\cdots,
\eeq
where $h$ is the lowest dimension in the operator spectrum of the CFT, and $\mathcal N$ is the multiplicity of operators with dimension $h$.  These R\'enyi entropies were analytically continued to $n=1$ in \cite{Calabrese:2010he}, giving the entanglement entropy
 \beq\label{s11x}
 S = - \mathcal N \(\frac{x}{4}\)^{2 h} \frac{\sqrt{\pi}}{4} \frac{\Gamma(2 h +1)}{\Gamma\(2 h +\frac32\)}+\cdots.
 \eeq
Noting that this result is not multiplied by a factor of the central charge, we can see that it will appear at subleading order in the bulk semiclassical expansion. We will reproduce precisely (\ref{s11x}) from the bulk determinant, as well as obtain contributions that are higher order in $x$.

For the R\'enyi entropies there are additional universal terms in the small $x$ expansion that are multiplied by a factor of the central charge \cite{Calabrese:2010he}. We shall reproduce these terms from a classical bulk computation in section \ref{sec:twoloop}. These terms however, vanish upon taking the $n \to 1$ limit and therefore do not appear in the leading order entanglement entropy.

\subsection{Solving the differential equation}

The first step is to set up a systematic procedure to solve the monodromy problem in a small $x$ expansion. It was shown in \cite{Faulkner:2013yia} -- and we will recall in section \ref{sec:twoloop} below -- that in this limit the dominant bulk saddle is given by the Schottky uniformization in which the monodromy around the $[-y,y]$ cycle is trivial. We must therefore solve for the accessory parameters that trivialize this monodromy and, with these parameters, also find the nontrivial monodromy around the remaining cycle, enclosing $[-1,-y]$. We achieve both of these steps at once with the following method.

Consider the following ansatz for two independent solutions to the differential equation (\ref{eq:diffeq}) in the regime $|z| \ll 1$
\be\label{eq:zexpand}
\psi_2^\pm = (z+y)^{\Delta_\pm} (z-y)^{\Delta_\mp} \sum_{m=0}^\infty \psi_2^{\pm(m)}(y) \, z^m \,,
\ee
where we can always normalize the solutions so that $\psi_2^{\pm(0)} = 1$. Here $\Delta_\pm = \half(1\pm \frac{1}{n})$ in order to isolate the non-analytic behavior at the branch points. In a generic situation, the series expansion in $z$ in (\ref{eq:zexpand}) would have radius of convergence $y$ due to the singular points in the differential equation at $z = \pm y$. However, the condition of trivial monodromy around the $[-y,y]$ cycle is seen to be precisely the condition that, after stripping off the prefactor in (\ref{eq:zexpand}), there is no further singularity at $z = \pm y$. If we can find accessory parameters such that this is true, then we can expand
\be\label{eq:yexpand}
\psi_2^{\pm(m)}(y) = \sum_{k=0}^\infty \psi_2^{\pm(m,k)} y^k \,,
\ee
with no negative powers of $y$. If this second expansion is possible, then the radius of convergence for the $z$ expansion in (\ref{eq:zexpand}) will have increased from $y$ to $1$, as the closest singular points are now at $z = \pm 1$. We can further expand the remaining unfixed accessory parameter $\g_1$ (recall that three of the four parameters are fixed by the conditions (\ref{eq:conditions}) for trivial monodromy at infinity) in terms of nonnegative powers of $y$. We find that if we expand the entire differential equation in a double power series expansion in $z$ and $y$ and demand the absence of negative powers of $y$, then we can uniquely solve the equation for the coefficients $\psi_2^{\pm(m,k)}$ and the accessory parameter $\g_1$. To the lowest few orders the accessory parameter is found to be
\begin{align}
\gamma_1 &= \left(\frac{1}{2}-\frac{1}{2 n^2}\right)+\frac{2 \left(n^2-1\right)^2 y^2}{3 n^4}+\frac{2 \left(n^2-1\right)^2 \left(49 n^4-2 n^2-11\right) y^4}{135 n^8}+\ocal\left(y^6\right)\,, \label{eq:gamma1}
\end{align}
and the solution is
\begin{align}
\psi_2^+ &= (z+y)^{\Delta_+}(z-y)^{\Delta_-} \[1+z \left(-\frac{(n^2-1) y}{3 n^3}+\frac{(-13 n^6+6 n^4+18 n^2-11) y^3}{135 n^7}+\ocal(y^5)\right)\right. \nonumber \\
&\qquad +z^2\left(\frac{1}{6} \left(\frac{1}{n^2}-1\right)+\frac{\left(4 n^6+3 n^4-9 n^2+2\right) y^2}{135 n^6}+\ocal\left(y^4\right)\right)  \\
&\qquad\left. +z^3\left(-\frac{\left(n^4-1\right) y}{30 n^5}+\ocal\left(y^3\right)\right)+z^4 \left(\frac{-11 n^4+10 n^2+1}{120 n^4}+\ocal\left(y^2\right)\right)+\ocal\left(z^5\right)\]. \nonumber
\end{align}
We have set up a recurrence relation that allows us to easily find these expansions to high order. Given $\psi_2^+$, the other solution $\psi_2^-$ is found by letting $y \to -y$.

The solutions $\psi_2^\pm$ are not sufficient to find the monodromy around the remaining $[-1,-y]$ cycle, as they are only valid for $|z| \ll 1$. In order to compute the monodromy we need, in addition, a second pair of solutions $\psi_1^\pm$ that hold for $y \ll |z|$. Because $y$ is small, these two pairs of solutions will have a parametric regime of overlap. By matching them in the overlap regime we will be able to find the monodromy. Fortunately, it turns out we can obtain $\psi_1^\pm$ with almost no effort! With the four singular points chosen to be at $\pm 1$ and $\pm y$, the differential equation (\ref{eq:diffeq}) is in fact invariant under the inversion $z \to y/z$ together with taking $\psi \to (z/y) \psi$. Therefore, given the solution we have obtained for $\psi_2^\pm$, valid at $|z| \ll 1$, using this map we immediately obtain solutions $\psi_1^\pm$, valid for 
$y \ll |z|$.

\subsection{Computing the monodromies}

With the solution to the differential equation in overlapping regimes at hand, we proceed to obtain the monodromy matrices $L_i$. As we discussed around equation (\ref{eq:othergen}) above, these can be built out of the monodromies $M_1$ and $M_2$ obtained by encircling the branch points $z=1$ and $z=y$, respectively. Near the $z=1$ branch point we use the $\psi_1^\pm$ solutions while near the $z=y$ branch point we can use the $\psi_2^\pm$ solutions. The relevant leading order behaviors near the branch points are
\be
\psi_1^\pm=(1+z)^{\Delta_\pm}(1-z)^{\Delta_\mp},\qquad
\psi_2^\pm=(z+y)^{\Delta_\pm}(z-y)^{\Delta_\mp} \,.
\ee
From these expression it is easy to compute $M_1$ and $M_2$. However, they will be expressed in different bases. To obtain the monodromies $L_i$ we need to relate these bases. We will work in the basis $(\psi_2^+,\psi_2^-)$, in which $M_2$ becomes diagonal:
\beq
M_2=\begin{pmatrix}e^{2\pi i\Delta_{+}} & 0 \\
0 & e^{2\pi i\Delta_{-}}
\end{pmatrix}.
\eeq
In this basis $M_1$ is non-diagonal. Let us write
\be
M_1=T^{-1}M_2T \,,
\ee
where $T$ is the transformation matrix between the two sets of bases:
\beq
\begin{pmatrix}\psi_1^+ \\
\psi_1^- \\
\end{pmatrix}=T\begin{pmatrix}\psi_2^+ \\
\psi_2^- \\
\end{pmatrix} \,.
\eeq
We can systematically find $T$ from this last equation by expanding $\psi_1^\pm$ and $\psi_2^\pm$, as obtained in the previous subsection, in the overlapping regime $y \ll |z| \ll 1$ and matching. To the lowest few orders this gives
\bea
\lefteqn{T_{11} = \frac{n}{2 y}+\frac{1}{2 n}-\frac{\left(n^4+n^2-2\right) y}{18 n^3}+\frac{\left(n^4+n^2-2\right) y^2}{18 n^5}} \nonumber\\
&& -\frac{\left(76 n^8+80 n^6-102 n^4-205 n^2+151\right) y^3}{4050 n^7}+\ocal\left(y^4\right),\\
T_{12} & = & T_{11}|_{y\to-y},\qquad
T_{21}=-T_{11}|_{y\to-y},\qquad
T_{22}=-T_{11},\qquad
T^{-1}=y \, T \,.
\eea
Note in particular the simple expression for $T^{-1}$ in terms of $T$. Thus, from (\ref{eq:othergen}) and the explicit expressions above for $M_2$ and $T$, we have explicit expressions for the Schottky generators $L_i$:
\be\label{eq:LT}
L_i = M_2^i \, T^{-1} M_2 T M_2^{-(i-1)} \,, \qquad i = 1, \ldots, n-1 \,.
\ee

\subsection{Summing over words and leading order result}
\label{sec:cdw}

The next step is to form all possible primitive (non-repeated) words from the Schottky generators $L_i$ and their inverses, up to conjugation in the Schottky group $\Gamma$. We must then compute the eigenvalues of these words and use these eigenvalues to evaluate the one-loop determinant contributions to the entanglement entropies given in section \ref{sec:oneloop}.

There are infinitely many such primitive conjugacy classes for $n>2$. However, we will shortly show that only finitely many of them contribute to the result at each order in a small $y$ expansion. In particular, at leading order in $y$, the only words that contribute will be seen to be built from ``consecutively decreasing'' $L_i$ (up to conjugation or inversion). These are words of the form 
\beq\label{eq:CDWone}
\gamma_{k,m} \equiv L_{k+m}L_{k+m-1}\cdots L_{m+1}.
\eeq
These consecutively decreasing words (CDWs) have the property that when written in terms of the $M_2$ and $T$ matrices they contain only one pair of $T$ and $T^{-1}$.  Explicitly, the CDW $\gamma_{k,m}$ can be written as
\beq\label{eq:cdw}
\gamma_{k,m}=M_2^{m+k} T^{-1} M_2^k T M_2^{-m}.
\eeq
It is immediate from the above expression that the eigenvalues of $\gamma_{k,m}$ only depend on its length $k$. Furthermore, the larger (in absolute value) eigenvalue is of order $1/y$ since $T$ is of order $1/y$ and $T^{-1}$ is of order 1. Specifically, to lowest order:
\beq\label{eq:tlead}
T=\frac{n}{2y}\begin{pmatrix}1 & -1 \\
1 & -1 \\
\end{pmatrix},\qquad
T^{-1}=yT\,.
\eeq

All words that are not related to any CDW by conjugation or inversion come with at least two pairs of $T$ and $T^{-1}$, being of the general form
\beq
\gamma_{k_1,k_2,\cdots,k_{2p},m} \equiv M_2^{m} \(\prod_{j=1}^p M_2^{k_{2j-1}} T^{-1} M_2^{k_{2j}} T \) M_2^{-m}.
\eeq
Their larger eigenvalues are of order $y^{-p}$ where $p\ge2$ is the number of pairs of $T$ and $T^{-1}$. We now check that the leading $y^{-p}$ term cannot vanish. We can calculate the large eigenvalue to leading order using only the leading order matrices (\ref{eq:tlead}). This gives
\beq
M_2^{-m} \gamma_{k_1,k_2,\cdots,k_{2p},m} M_2^{m} =\(\frac{n^2}{4y}\)^{p}
\begin{pmatrix}e^{2\pi ik_1\Delta_+} & -e^{2\pi ik_1\Delta_+} \\
e^{2\pi ik_1\Delta_-} & -e^{2\pi ik_1\Delta_-} \\
\end{pmatrix}
\prod_{j=2}^{2p} \(e^{2\pi ik_j\Delta_+}-e^{2\pi ik_j\Delta_-}\).
\eeq
The eigenvalues of this matrix can easily be calculated.  One of them is zero and the other gives us the large eigenvalue to leading order in small $y$:
\beq\label{qgamma}
q_\gamma^{-1/2}=\(\frac{n^2}{4y}\)^{p} \prod_{j=1}^{2p} \(e^{2\pi ik_j\Delta_+}-e^{2\pi ik_j\Delta_-}\)
=\(-\frac{n^2}{y}\)^{p} \prod_{j=1}^{2p} \sin(\pi k_j/n) \,.
\eeq
This term cannot be zero unless one of the $k_j$ is zero, which means that we should have contracted a neighboring pair of $T$ and $T^{-1}$ and decreased $p$ by 1.

Recalling the formulae (\ref{eq:zoneloop}) and (\ref{eq:scalarloop}) for the one-loop contribution to the higher genus partition functions, we conclude that only CDWs and their inverses contribute to leading order in the small $y$ limit. This is because our result above implies that the smaller of the eigenvalues, $q_\gamma$, vanishes like $y^p$. Therefore the largest contribution to (\ref{eq:zoneloop}) or (\ref{eq:scalarloop}) at small $y$ comes from $p=1$. For the CDW $\gamma_{k,m}$ we can apply \eqref{qgamma} with $k_1=k_2=k$ and obtain
\beq\label{eq:qgamma}
q_\gamma^{-1/2}= -\frac{n^2}{y} \sin^2(\pi k/n) +\ocal(1).
\eeq

Given the eigenvalues (\ref{eq:qgamma}) we are finally ready to evaluate the determinants. Firstly, note that there are $n-k$ inequivalent CDWs of length $k$ for each $1\le k\le n-1$, labeled by $0\le m\le n-k-1$ in the $\gamma_{k,m}$ given in equation (\ref{eq:cdw}). Doing the graviton case first, the contribution to the sum in (\ref{eq:zoneloop}) from these CDWs and their inverses gives, using the relation (\ref{eq:sz}) between the higher genus partition functions and the R\'enyi entropies,
\begin{align}
\left. S_n \right|_\text{one-loop}=& - \frac{1}{n-1}\sum_{\gamma\in\mathcal P} \Real \[q_\gamma^2+\ocal(q_\gamma^3)\]
= - \frac{2}{n-1}\(\frac{y}{n^2}\)^4 \sum_{k=1}^{n-1}\frac{n-k}{\sin^8(\pi k/n)}+\ocal(y^5) \nonumber \\
=& - \frac{n}{n-1}\(\frac{x}{4n^2}\)^4 \sum_{k=1}^{n-1}\frac{1}{\sin^8(\pi k/n)}+\ocal(x^5), \label{eq:snoneloop}
\end{align}
where in the final step we have switched to the cross-ratio $x=4y+\ocal(y^2)$.

This result agrees with the general expression \eqref{s1nx} exactly, with $h=2$ corresponding to the stress tensor in 1+1 dimensions and with multiplicity $\mathcal N=2$.  Using the same analytic continuation we have therefore also reproduced the entanglement entropy \eqref{s11x} at leading order in the cross-ratio $x$.

We can furthermore consider the one-loop contribution of a bulk scalar with general scaling dimension $h$. Using the determinant formula (\ref{eq:scalarloop}) and following the logic above, we reproduce precisely the leading order small $x$ result \eqref{s11x}, now at general $h$.

\subsection{Higher orders in the small $x$ expansion}

Having successfully reproduced the universal leading order term \eqref{s11x} in the entanglement entropy from the bulk, we can now systematically extend the method above to compute the entanglement entropy to high order in an expansion in small cross-ratio $x$. For concreteness we will focus on the graviton case.
The expansion involves firstly finding and matching the power series solutions to the differential equation to higher order, allowing construction of the Schottky generators (\ref{eq:LT}) to high order. This can be completely systematized via a recursion relation. Second, we must include words that are not CDWs in the sum over primitive conjugacy classes ${\mathcal P}$ in the determinant formulae (\ref{eq:zoneloop}). This can also be done completely systematically: we saw in (\ref{eq:qgamma}) and (\ref{eq:snoneloop}) that at order $y^{4p}$ in the entropies we need only include words that can be written in terms of at most $p$ pairs of $T$ and $T^{-1}$. These words are formed by joining at most $p$ CDWs (or their inverses), and we will refer to them as $p$-CDWs in the following. At higher order we will also need to keep an increasing number of terms in the product in (\ref{eq:zoneloop}).

The strategy in the paragraph above will give us the R\'enyi entropies. The technically most challenging part is the analytic continuation of these expressions to $n=1$, in order to obtain the entanglement entropy itself. The computation can be organized in terms of the complexity of the words contributing to the determinant formulae. We can start with the CDW contribution. Computing the eigenvalues of the CDWs to high order in $y$ and plugging them into the determinant (\ref{eq:zoneloop}) we find the contribution to the R\'enyi entropies
\bea
\lefteqn{S_{n,\text{CDW}} = - \frac{n}{n-1}\sum_{k=1}^{n-1} \left[ \frac{c^8 x^4}{256 n^8}+\frac{x^5 \left(c^{10}+c^8 n^2-c^8\right)}{128 n^{10}} \right.} \label{eq:sncdw} \\
&& \left.+\frac{x^6}{36864 n^{12}} \left(405 c^{12}+720 c^{10} n^2-720 c^{10} 
 +404 c^8 n^4-712 c^8 n^2+308 c^8\right) + \ocal(x^7)\right] \,. \nonumber
\eea
Here $c \equiv \text{csc}(\pi k/n)$. We have computed this expansion to order $x^{12}$ and there is no difficulty in principle in working to very high order. The analytic continuation of this expression is achieved using the result from \cite{Calabrese:2010he} that
\be\label{eq:continue}
\lim_{n \to 1} \frac{1}{n-1} \sum_{k=1}^{n-1} \left[ \csc (\pi k/n) \right]^{2 \a} = \frac{\G(3/2) \G(\a + 1)}{\G(\a + 3/2)} \,.
\ee
Using this result to analytically continue the expansion (\ref{eq:sncdw}), we obtain the contribution to the entanglement entropy
\bea
\lefteqn{S_\text{CDW} = - \left(\frac{x^4}{630} + \frac{2 x^5}{693} + \frac{15 x^6}{4004} + \frac{x^7}{234} \right.} \label{eq:scdw} \\
&& \left.  + \frac{2003 x^8}{437580} + \frac{1094 x^9}{230945} + \frac{4660 x^{10}}{969969} + \frac{9760 x^{11}}{2028117} + \frac{1939877 x^{12}}{405623400} + \ocal(x^{13}) \right) \,. \nonumber
\eea
Perhaps in the future this function will be obtained in closed form. Because the non-CDW contribution to the entropies starts at order $x^8$, as we discussed above, the above expression gives the exact entanglement entropy up to order $x^7$.

The analytic continuation of the non-CDW contributions to the R\'enyi entropies becomes progressively difficult as the complexity of the words increases. A systematic treatment is possible, as we now illustrate by computing the 2-CDW contribution to the entanglement entropy at order $x^8$. This will give us the following full one-loop contribution to the entanglement entropy up to this order
\be
\left. S \right|_\text{one-loop}=  - \left(\frac{x^4}{630} + \frac{2 x^5}{693} + \frac{15 x^6}{4004} + \frac{x^7}{234} + \frac{167 x^8}{36936} + \ocal(x^{9}) \right) \,.\label{eq:fulls}
\ee
We see that adding the CDW and 2-CDW contributions has simplified $x^8$ term a little relative to (\ref{eq:scdw}).

From our discussion above, the first non-CDW corrections come from (primitive) words which can be built from 2-CDWs.  To classify the 2-CDWs, it is convenient to first re-organize our notation for CDWs a little. Let us write the CDWs and their inverses as
\beq
\gamma_{[m_1,m_2]}= M_2^{m_1} T^{-1} M_2^{m_1-m_2} T M_2^{-m_2}=
\begin{cases}
L_{m_1} L_{m_1-1} \cdots L_{m_2+1},& \text{if } m_1>m_2\\
\gamma_{[m_2,m_1]}^{-1},& \text{if } m_2>m_1
\end{cases} \,,
\eeq
where $m_1$ and $m_2$ take integer values from 0 to $n-1$. The square brackets are to remind us that the indices labeling the CDWs have a different meaning to those in (\ref{eq:CDWone}). From now on we will use the term CDW to refer to either a CDW or its inverse, namely $\gamma_{[m_1,m_2]}$ for any $m_1\neq m_2$.  With this notation, all 2-CDWs may be written as
\beq\label{2cdw}
\gamma_{[m_1,m_2]}\gamma_{[m_3,m_4]} = M_2^{m_1} T^{-1} M_2^{m_1-m_2} T M_2^{-m_2+m_3} T^{-1} M_2^{m_3-m_4} T M_2^{-m_4} \,,
\eeq
where $m_1\neq m_2$, $m_3\neq m_4$, $m_2\neq m_3$, $m_4\neq m_1$. The last two conditions ensure that the two CDWs do not join into a CDW.  Since we need to sum over primitive conjugacy classes, we impose $(m_1,m_2)\neq(m_3,m_4)$ to ensure that \eqref{2cdw} is primitive. Finally, noting that exchanging $(m_1,m_2)$ with $(m_3,m_4)$ merely conjugates the word, we have therefore overcounted and should divide the final sum by 2.

The eigenvalues of the word \eqref{2cdw} were worked out in \eqref{qgamma} to leading order in $y$. The larger eigenvalue is given by
\beq
\textstyle q_{\gamma}^{-1/2}=\(\frac{n^2}{y}\)^{2} \sin\(\frac{\pi(m_1-m_2)}{n}\) \sin\(\frac{\pi(m_2-m_3)}{n}\) \sin\(\frac{\pi(m_3-m_4)}{n}\) \sin\(\frac{\pi(m_4-m_1)}{n}\).
\eeq
It follows that the leading order contributions from all 2-CDWs to the R\'enyi entropy at small cross-ratio is
\bea\label{sn2}
\lefteqn{S_{n,2-\text{CDW}}=} \\
&& - \frac{1}{n-1} \(\frac{y}{n^2}\)^8 \frac{1}{2} \sum_{\{m_j\}} \frac{1}{\sin^4\(\frac{\pi(m_1-m_2)}{n}\) \sin^4\(\frac{\pi(m_2-m_3)}{n}\) \sin^4\(\frac{\pi(m_3-m_4)}{n}\) \sin^4\(\frac{\pi(m_4-m_1)}{n}\)}, \nonumber
\eea
where the range of the sum is defined as
\bea\label{sn2r}
& 0\le m_1,m_2,m_3,m_4 \le n-1,\, & \nonumber \\
& m_1\neq m_2,\, m_3\neq m_4,\, m_2\neq m_3,\, m_4\neq m_1,\,
(m_1,m_2)\neq(m_3,m_4) \, .&
\eea
In appendix \ref{sec:2cdw} we evaluate this sum in closed form for integer $n$. The answer is
\begin{multline}\label{eq:sn2cdw}
S_{n,2-\text{CDW}}= - \(\frac{y}{n^2}\)^8\frac{2 (n-2) n (n+1) (n+2)}{488462349375}\left(5703 n^{12}+192735 n^{10}+3812146 n^8\right.\\
\left.+75493430 n^6+1249638099 n^4+9895897835 n^2-162763727948\right).
\end{multline}
It is now very simple to analytically continue this result to $n \to 1$. The answer for the entanglement entropy is
\beq
S_{2-\text{CDW}} = - \frac{29}{510510} x^8 + \ocal(x^9).
\eeq
Adding this correction to the CDW contribution (\ref{eq:scdw}) we obtain the previously advertised expression (\ref{eq:fulls}) for the full one-loop entanglement entropy up to order $x^8$.

While we could proceed systematically to higher orders, the steps become increasingly cumbersome. A more efficient approach is desirable.

\section{Two intervals on a line: one-loop R\'enyi entropies}
\label{sec:twoloop}

In the previous section we found, as is commonly the case, that the most challenging step in computing the entanglement entropy is the analytic continuation of the R\'enyi entropies. However, the R\'enyi entropies themselves carry information about the entanglement structure of the theory and are much easier to compute. In this section we present numerical results for the one-loop contribution to the R\'enyi entropies, as well as an exact result for the simplest case of $S_2$. One upshot of this section will be that our previous analytic expansion of the one-loop R\'enyi and entanglement entropies to order $x^8$ leads to a numerically accurate computation of these entropies at all values of the cross-ratio $x$.

\subsection{Exact result for $S_2$}

For $n=2$, the differential equation (\ref{eq:diffeq}) for the Schottky uniformization map can be solved exactly \cite{Faulkner:2013yia}, see also \cite{Lunin:2000yv, Headrick:2010zt} for explicit uniformizations in this case. The branched cover has genus one, and correspondingly the solution is given in terms of elliptic integrals. In particular, the map (\ref{eq:wdef}) is \cite{Faulkner:2013yia}
\be\label{eq:wsol}
w = e^{2 \, h \, t(z)} \,, \qquad t'(z) = \frac{1}{\sqrt{z(z-1)(z-x)}} \,.
\ee
Here $x$ is the cross-ratio (\ref{eq:cross}) and $h$ is a constant determined by the accessory parameter. Here we are taking the intervals to be $[0,x]$ and $[1,\infty]$. From (\ref{eq:wsol}), the monodromy around a closed cycle is
\be
w \mapsto w \, e^{2 h \oint t'(z) dz} \,.
\ee
Requiring trivial monodromy around the $[0,x]$ cycle then imposes
\be
h = - \frac{i \pi}{4 K(x)} \,.
\ee
The remaining nontrivial monodromy around a cycle enclosing $x$ and $1$ generates the Schottky group. The generator is found to be
\be\label{eq:unique}
w \mapsto L(w) = w \, e^{- 2 \pi K(1-x)/K(x)} \,.
\ee

The unique generator (\ref{eq:unique}) of the Schottky group in this case has eigenvalue
\be
q = e^{- 2 \pi K(1-x)/K(x)} \equiv e^{2 \pi i \tau}\,.
\ee
We introduced here the modular parameter $\tau$ of the torus.
Using this eigenvalue in the determinant formula (\ref{eq:zoneloop}) gives the one-loop contribution to the R\'enyi entropy
\be\label{eq:twoexact}
\left. S_2 \right|_\text{one-loop} = 2 \sum_{m=2}^\infty \log(1 - q^m)  = 2 \log \left(\frac{q^{-\frac{1}{24}} \eta(\tau)}{1 - q} \right) \,.
\ee
This result can also be stated in terms of theta functions or $q$-Pochhammer symbols. Expanding the exact result (\ref{eq:twoexact}) in small cross-ratio $x$ we obtain
\be
\left. S_2 \right|_\text{one-loop} = - \left(\frac{x^4}{32768} +\frac{x^5}{16384} + \frac{721 x^6}{8388608} + \frac{883 x^7}{8388608} + \frac{515395 x^8}{4294967296} + \mathcal O (x^9) \right) \,.
\ee
This expansion is seen to agree exactly with the expansion (\ref{eq:sncdw}) when $n=2$. This provides a nice check of our more general expansion of the R\'enyi entropies.

\subsection{Numerical results for higher R\'enyi entropies}

We describe how to calculate numerically both the classical and the 1-loop contributions to the holographic R\'enyi entropies for two disjoint intervals in a 1+1 dimensional CFT.

\subsubsection{Classical contribution}

At the classical level, the R\'enyi entropies are given in terms of the accessory parameters
\beq\label{snclass}
\frac{\partial S_{n}}{\partial z_i}=-\frac{cn}{6(n-1)} \gamma_i \,,
\eeq
as we reviewed in section \ref{sec:classical} above. The accessory parameters are determined by requiring trivial monodromy around either the $[z_1,z_2]$ or the $[z_2,z_3]$ cycle, as we have discussed. The $\gamma_i$ determined in this way are functions of $n$ and all four $z_i$.  We may then integrate \eqref{snclass} to get $S_{n}$.  We fix the integration constant by requiring that when one of the intervals shrinks to zero size, the R\'enyi entropy should be equal to the R\'enyi entropy of a single interval, which is given for all CFTs by \cite{Holzhey:1994we}
\beq\label{Snone}
S_{n}(L)= \frac{c}{6} \(1+\frac{1}{n}\) \log\frac{L}{\epsilon} \qquad\text{(for a single interval)} \,.
\eeq
Here $L$ is the length of the interval and $\epsilon$ a short distance cutoff.

We must compare the two values for $S_{n}$ obtained by trivializing the monodromy on either of the two cycles, and choose the smaller one. This corresponds to taking the dominant saddle point in the bulk path integral.

It is useful to consider the mutual R\'enyi information $I_n$ defined as
\beq\label{In}
I_n (L_1:L_2) = S_n(L_1) + S_n(L_2) - S_n(L_1 \cup L_2)
\eeq
where by a slight abuse of notation we have denoted the two disjoint intervals by $L_1$ and $L_2$, and $S_n(L_1 \cup L_2)$ is the R\'enyi entropy of their union (which we have been calling $S_n$).  The mutual R\'enyi information $I_n$ is cutoff-independent, and has the nice property that in a CFT it depends on the four $z_i$ only through the cross-ratio $x$. It furthermore satisfies (see e.g. \cite{Headrick:2010zt})
\beq\label{Isym}
I_n(1-x)= I_n(x) + \frac{c}{6}\(1+\frac{1}{n}\) \log\frac{1-x}{x}.
\eeq
This relation is exact in an arbitrary CFT.  At the classical level in our holographic context, it will allow us to avoid having to calculate the other saddle point when we cross a phase transition at $x=1/2$.

\begin{figure}[h]
        \centering
        \includegraphics[width=0.49\textwidth]{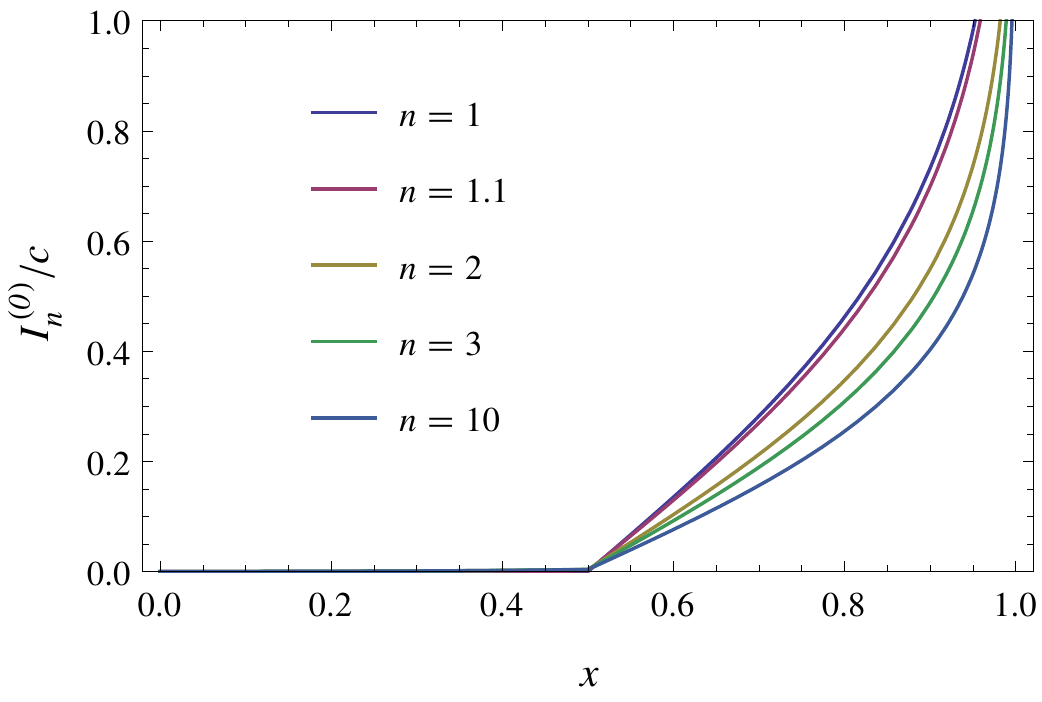}
        \includegraphics[width=0.50\textwidth]{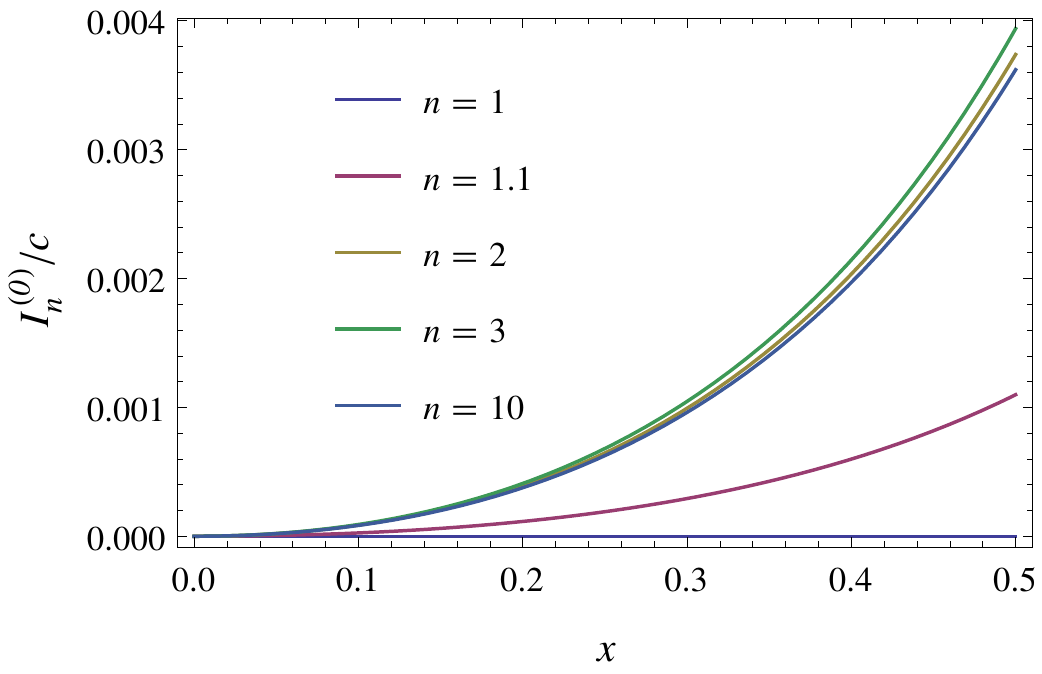}
\caption{The mutual R\'enyi information $I_{n}^{(0)}$ of two disjoint intervals, at the classical level, plotted as functions of the cross-ratio $x$ for various $n$.  The plot on the right is the same as the one on the left except that it is restricted to $x<1/2$, showing more clearly that the mutual R\'enyi informations with $n>1$ do not vanish for $x < 1/2$. \label{fig:I0}}
\end{figure}

In Fig.\ \ref{fig:I0} we show the mutual R\'enyi information $I_n$ at the classical level as a function of the cross-ratio $x$ for various $n$.  Note that the procedure outlined above makes sense for real $n$ (not just integers).  The $n=1$ curve corresponding to the mutual information is obtained by taking the $n\to1$ limit, either numerically or analytically.  The analytic result obtained in \cite{Faulkner:2013yia} for $n=1$ is
\beq
I_1=
\begin{cases}
0&\quad 0 \le x \le 1/2 \\
\displaystyle \frac{c}{3}\log\frac{1-x}{x}&\quad 1/2 \le x < 1
\end{cases} \,,
\eeq
which satisfies \eqref{Isym}. The fact that the mutual information vanishes identically for $x \leq 1/2$ is a distinctive feature of the leading order holographic entanglement. Therefore at small $x$, the one-loop entanglement we have computed in section \ref{sec:univ} is the leading nontrivial contribution. This particular behavior of the multi-interval holographic entanglement is closely tied to the behavior we emphasized in section \ref{sec:classical} above for the entanglement entropy of a single interval on a circle at finite temperature. This is because, as pointed out in \cite{Headrick:2010zt}, the phase transition between saddles trivializing the monodromy around the $[z_1,z_2]$ and $[z_2,z_3]$ cycles, respectively, is a close cousin of the Hawking-Page transition.

As shown in the second plot of Fig.\ \ref{fig:I0}, the mutual R\'enyi entropies with $n>1$ do not vanish identically for $x < 1/2$. In fact, from our previous small $x$ result for the accessory parameters (\ref{eq:gamma1}) and the general formulae (\ref{snclass}) and (\ref{In}) for the classical R\'enyi mutual information we obtain
\be\label{eq:tworn}
I_n = \frac{c (n-1) (n+1)^2}{144 n^3} x^2 + \cdots \,.
\ee
This agrees with our numerical results and also with a universal contribution to the small $x$ R\'enyi entropy from the energy momentum tensor that was found in \cite{Calabrese:2010he}. We see that this contribution vanishes in the $n \to 1$ limit and therefore does not appear in the entanglement entropy.
A similar fact will be observed in the following section at finite temperature. It is specifically in the $n \to 1$ entanglement entropy limit that a certain amount of structure is erased from the leading order holographic entanglement.

\subsubsection{One-loop contribution}

In this subsection we present numerical results on one-loop R\'enyi entropies for two disjoint intervals. Our basic strategy is
\begin{enumerate}
\item Find the correct accessory parameters by imposing trivial monodromy on either the $[z_1,z_2]$ or the $[z_2,z_3]$ cycles. This step is already done in computing the entanglement at the classical bulk level.
\item Find the remaining nontrivial monodromy matrices. This involves a set of $n-1$ independent cycles, giving the $n-1$ generators $L_i$ of the Schottky group.  We need to be careful in choosing the same basis for all monodromy matrices.
\item Form primitive words $\gamma$ up to conjugation within the Schottky group.  For $n\ge3$ there are infinitely many such words, so we impose a cutoff on the word length in order to sum over their contributions to the one-loop R\'enyi entropies.
\item Evaluate the two eigenvalues of each primitive conjugacy class $\gamma$ and find the smaller (in magnitude) eigenvalue $q_{\gamma}^{1/2}$. We then sum over their contributions by rewriting the determinant \eqref{eq:zoneloop} in terms of the Dedekind eta function:
\beq
\left. S_{n} \right|_\text{one-loop}= \frac{1}{n-1} \sum_{\gamma\in\mathcal P} \log \left|\frac{q_{\gamma}^{-1/24}\eta(q_{\gamma})} {1-q_{\gamma}}\right|.
\eeq
\end{enumerate}

The one-loop R\'enyi entropy $S_{n}$ obtained in this way is always negative.  We can also consider the one-loop mutual R\'enyi information
\be
\left. I_{n} \right|_\text{one-loop} = - \left. S_{n} \right|_\text{one-loop} \,. 
\ee
This is apparent from the definition of the mutual informations \eqref{In} and the fact that the single interval result \eqref{Snone} is exact (i.e.\ has no one-loop correction). It follows that, as we mentioned above, the one-loop R\'enyi entropies are also only a function of the cross-ratio $x$. It also follows that the one-loop R\'enyi entropies will be symmetric under $x \to 1- x$.

\begin{figure}[h]
\centering
\includegraphics[width=0.49\textwidth]{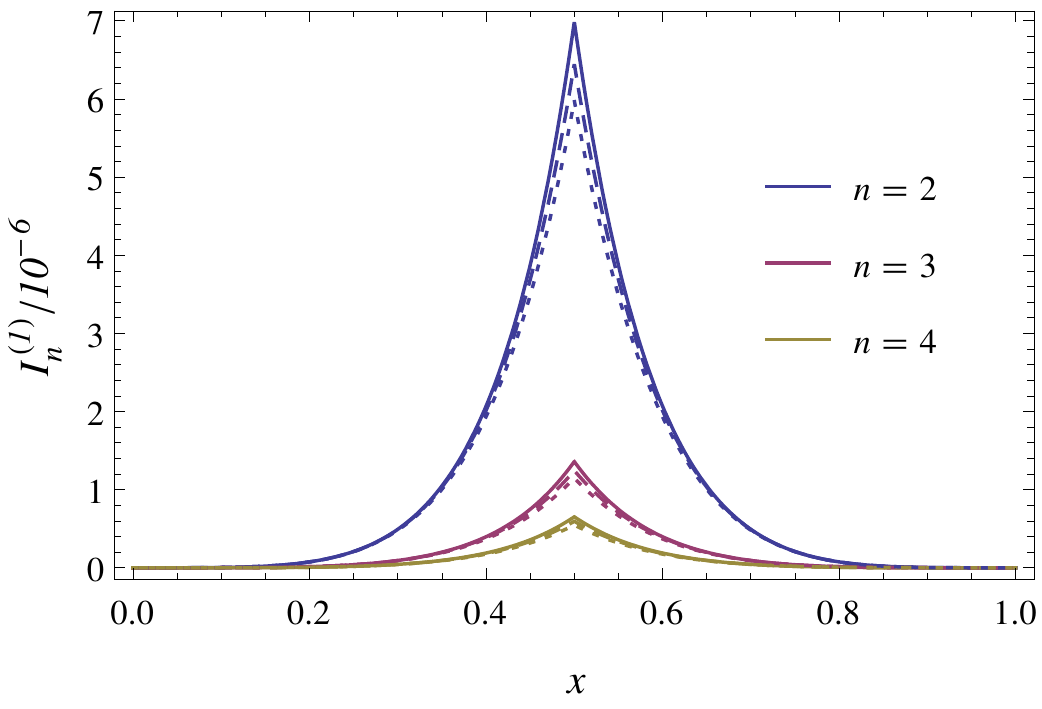}
\includegraphics[width=0.49\textwidth]{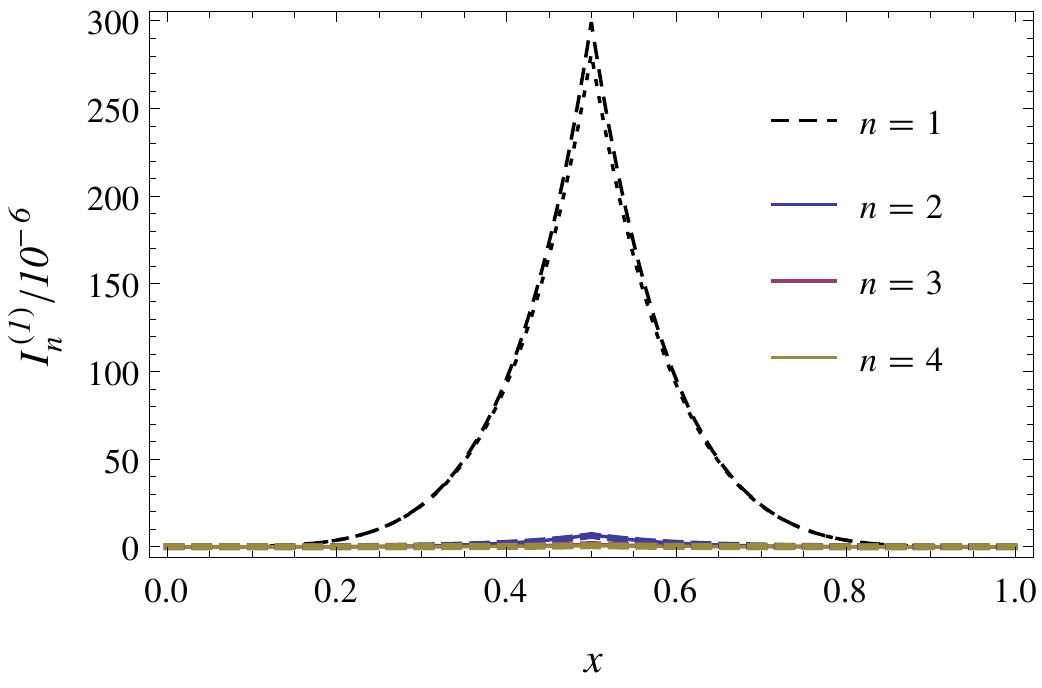}     
\caption{\label{fig:I1}(a) The one-loop contribution to the mutual R\'enyi information $I^{(1)}_{n}$ of two disjoint intervals plotted as functions of the cross-ratio $x$ for various $n$.  The solid curves come from numerical calculations, and the dashed (or dotted) curves come from an analytic expansion in $x$ up to $x^8$ (or $x^7$) inclusive.  (b) The same plot but now with $n=1$ black dashed/dotted curves from the analytic expansion.}
\end{figure}

In Fig.\ \ref{fig:I1} we show both numerical and analytical results for the one-loop mutual R\'enyi information $I^{(1)}_{n}$ for various $n$.  The numerical results are shown in solid curves and are obtained by including words of length 4 or smaller.  We find numerically that contributions of all words of length $k$ decrease very fast with increasing $k$ for $k>n-1$.\footnote{We know from the analytic expansion of section \ref{sec:univ} that for small $x$ the leading contributions come from CDWs whose lengths range from 1 to $n-1$, so we have to include at least words of length $n-1$ or smaller to capture the leading order small $x$ contributions.}  In fact with the choice of cutting off the length at $k_{\max}=4$ we can estimate a conservative upper bound for the error $\Delta I^{(1)}_{n}$ from contribution we get from words exactly of length 4.  We find that the upper bound for the relative error $\Delta I^{(1)}_{n} / I^{(1)}_{n}$ is extremely small, of order $10^{-13}$ for $n=3$ and of order $10^{-7}$ for $n=4$. For $n=2$ there are no primitive words beyond the generators $L_i$ and their inverses, so we have not neglected any contribution.

The analytic results are shown in Fig.\ \ref{fig:I1} in dashed and dotted curves.  The dashed curves come from the small $x$ expansion up to $x^8$ inclusive. This computation was described in section \ref{sec:univ}. For reference we quote here the complete expression
\bea 
\lefteqn{\left. I_n \right|_\text{one-loop} =  \frac{(n+1) \left(n^2+11\right) \left(3 n^4+10 n^2+227\right) x^4}{3628800 n^7} } \\
&& +\frac{(n+1) \left(109 n^8+1495 n^6+11307 n^4+81905 n^2-8416\right) x^5}{59875200 n^9} \nonumber\\
&& +\frac{(n+1) x^6}{523069747200 n^{11}} \left(1444050 n^{10}+19112974 n^8+140565305 n^6+1000527837 n^4\right.\nonumber\\
&& \left.-167731255 n^2-14142911\right)
+\frac{(n+1) x^7}{1569209241600 n^{13}} \left(5631890 n^{12}+72352658 n^{10}\right.\nonumber\\
&& \left.+520073477 n^8+3649714849 n^6-767668979 n^4-140870807 n^2+13778112\right)\nonumber\\
&& +\frac{(n+1) x^8}{3766102179840000 n^{15}} \left(16193555193 n^{14}+202784829113 n^{12}+1429840752361 n^{10}\right.\nonumber\\
&& \left.+9916221391201 n^8-2370325526301 n^6-689741905741 n^4+59604098747 n^2+161961045427\right).\nonumber
\eea
The dotted curves come from the same expansion but without the $x^8$ terms.  We see that the dashed curves are close to the solid ones (from the numerical study) even at $x=1/2$. The dotted curves are also very close, but including the $x^8$ terms made a visible difference and pushed the curves closer to the numerical result. This constitutes a check of our analytic expansion. The symmetry of the one-loop contribution under $x \to 1- x$ allows the small $x$ expansion result to give the mutual informations accurately over the whole range of $x$.

In the second plot of Fig.\ \ref{fig:I1} we added the $n=1$ results for the mutual information from the analytic expansion in small $x$.  They are shown as black dashed/dotted curves. Again the dashed curve is up to $x^8$ inclusive and the dotted curve is without the $x^8$ terms.  As we can see, the $n=1$ limit is much larger than the cases for $n\ge 2$.  We do not have a numerical curve for $n=1$, but we may estimate the error of the analytic expansion by taking the difference of the dashed and the dotted curves.  As we see this is rather small. It therefore seems that the second plot in Fig. \ref{fig:I1} gives an accurate description of the one-loop correction to the entanglement entropy at all cross-ratios $x$.

\section{One interval on a circle at high and low temperatures}
\label{sec:tloop}

In this section we calculate the R\'enyi and entanglement entropies for one interval in a CFT on a circle of length $R$ and in a thermal state with temperature $T$. By exploiting translational invariance we can choose the interval symmetrically as $[-y,y]$. Note that this is different from the choice made in section \ref{sec:tclassical} where the interval was $[y,R-y]$.

We will work analytically in either the high temperature or low temperature limit, and compare with known parametric results.  We will focus on the bulk one-loop contributions for the most part, but will also obtain the classical R\'enyi entropies (in a large- or small-$T$ expansion). Even though we will present results mostly at the leading order, our method allows for a systematic computation of subleading corrections. Our expansions will be sufficient to demonstrate the properties of thermal entanglement that were not visible at a classical order in the entanglement entropy.

\subsection{High temperature limit: systematic expansion}

Let us first focus on the high temperature limit.  By this we mean that $T$ is taken large in units of $1/R$ while we work with general $y$. We make no assumptions about $Ty$ being large or small.  As we will see, this means that we work in an expansion in $e^{-2\pi TR}$, and at each order we keep $Ty$ general.

In this subsection, we discuss how to perform the large-$T$ expansion systematically.  It is easiest to do this in a new coordinate $u$ defined by ``wrapping'' the time circle:
\beq
u \equiv e^{-2\pi T z} \,.
\eeq
Since the torus differential equation \eqref{eq:diffeqtorus} is periodic in time with period $1/T$, it is single-valued in the $u$ coordinate.  Furthermore, our desired solutions have trivial monodromy around the time circle at high temperatures, and therefore $u$ is also a natural coordinate in this regime.

Working in the complex $u$ coordinate, we can use \eqref{eq:wpsum} and \eqref{eq:zetasum} in the Appendix to rewrite the Weierstrass functions appearing in the torus differential equation \eqref{eq:diffeqtorus}:
\beq\label{eq:wpu}
\wp(z \pm y) = \sum_{m=-\infty}^{\infty} \frac{4\pi^2 T^2}{f(u u_y^{\pm1} u_R^m)} - \sum_{m\ne0} \frac{4\pi^2 T^2}{f(u_R^m)} +\frac{\pi^2 T^2}{3} \,,
\eeq
and similarly for $\zeta(z \pm y)$.  Here $f$ is defined as $f(u)=u+u^{-1}-2$, and $u_y$, $u_R$ are shorthand notations for
\beq
u_y \equiv e^{-2\pi T y} \,,\qquad
u_R \equiv e^{-2\pi T R} \,.
\eeq
From \eqref{eq:wpu} we see that the differential equation \eqref{eq:diffeqtorus} has a regular singular point at $u=u_y^{\pm 1} u_R^m$ for any integer $m$.

We will solve the differential equation \eqref{eq:diffeqtorus} in one spatial period $-R/2\le \Real z \le R/2$, which corresponds to an annular region $u_R^{1/2}\le |u| \le u_R^{-1/2}$ on the complex $u$ plane. Our strategy will be similar to that employed in section \ref{sec:univ} for the two interval case. This annular region contains two singular points $u=u_y^{\pm 1}$.  Therefore we consider the following ansatz for two independent solutions to the differential equation \eqref{eq:diffeqtorus}:
\beq\label{eq:uexpand}
\psi^{\pm} = \frac{1}{\sqrt u} (u-u_y)^{\Delta_\pm} \(u-\frac{1}{u_y}\)^{\Delta_\mp} \sum_{m=-\infty}^{\infty} \psi^{\pm(m)}(u_y,u_R)u^m \,,
\eeq
where we have isolated the non-analytic behavior at $u=u_y^{\pm 1}$ (or equivalently $z=\pm y$), similar to \eqref{eq:zexpand} in the two-interval case.  We have also pulled out a factor of $1/\sqrt{u}$, which gets a minus sign around the time circle (which is a circle centered at the origin in the complex $u$ plane).  This minus sign does not affect the monodromy in $PSL(2,\mathbb Z)$, and is the expected behavior from our small $n-1$ analysis\footnote{While varying $n$ continuously to general values, we do not expect a discontinuous change flipping the sign of the monodromy matrix (which is minus the identity matrix)around the time circle.} in section \ref{sec:tclassical}. As before, we define $\Delta_{\pm} = \frac12 (1\pm \frac{1}{n})$, and normalize the solutions so that $\psi^{\pm(0)}=1$.

The ansatz \eqref{eq:uexpand} is a Laurent expansion on the complex $u$ plane (after stripping off the prefactor), which has an inner and an outer circle of convergence.  For a generic accessory parameter $\gamma$ (and the constant $\delta$), the Laurent expansion in \eqref{eq:uexpand} is convergent only within a smaller annulus $u_y < |u| < u_y^{-1}$ due to the singular points at $u_y^{\pm 1}$.  Similarly to the discussions around \eqref{eq:yexpand}, the condition of trivial monodromy around the time circle (outside the interval $[-y,y]$) here is precisely the condition that, after stripping off the prefactor in \eqref{eq:uexpand}, the Laurent expansion itself is analytic at $u= u_y^{\pm1}$. This is because any branch cut not absorbed by the prefactor in \eqref{eq:uexpand} causes a nontrivial monodromy around the time circle.  If this is true, then the inner and outer circles of convergence have been pushed to $|u|=(u_R/u_y)^{\pm1}$, at which the next two singular points are found.  Expanding in small $u_R$ this means that each of the coefficients in \eqref{eq:uexpand} must be of the form
\beq\label{eq:uyexpand}
\psi^{\pm(m)}(u_y,u_R) = \sum_{k=|m|}^{\infty} \psi^{\pm(m,k)}(u_y) u_R^k \,.
\eeq
In a generic situation, the Taylor expansion \eqref{eq:uyexpand} would start at order $u_R^0$ instead of $u_R^{|m|}$, as the generic (inner and outer) radii of convergence are $u_y^{\pm1}$ which are of order $u_R^0$ in an expansion in $u_R$.  For the extended radii of convergence to be possible (giving trivial monodromy around the time circle), the accessory parameter $\gamma$ and the constant $\delta$ must take specific values.  By expanding the entire differential equation \eqref{eq:diffeqtorus} in $u$ and $u_R$, and demanding that \eqref{eq:uyexpand} be solutions, we can uniquely solve the coefficients $\psi^{\pm(m,k)}$ in \eqref{eq:uyexpand} as well as $\gamma$ and $\delta$, to any order in $u_R$ that we desire.  It is clear that we can keep $u_y$ (in other words $Ty$) general in this procedure.

To the lowest few orders the accessory parameter $\gamma$ and the constant $\delta$ are found to be
\begin{align}
&\begin{multlined}[0.9\textwidth]
\gamma = \pi T (1 + u_y^2) \[
\frac{1-n^2}{n^2 \left(u_y^2-1\right)}+\frac{\left(n^2-1\right)^2 \left(u_y^2-1\right)^3}{6 n^4 u_y^4} u_R^2 \right.\\
\left. +\frac{\left(n^2-1\right)^2 \left(u_y^2-1\right)^3 \left(3 u_y^4+2 u_y^2+3\right)}{12 n^4 u_y^6} u_R^3 +\ocal\left(u_R^4\right) \]\,,
\end{multlined} \nonumber\\\label{eq:tgamma}
&\begin{multlined}[0.9\textwidth]
\delta = \frac{\pi ^2 T^2}{n^2 \left(u_y^2-1\right)} \left\{
\frac{1}{6} \left[\left(n^2-1\right) \left(u_y^2+1\right) \log(u_y) -\left(7 n^2-1\right) \left(u_y^2-1\right)\right] \right.\\
\left.\phantom{\frac{1}{6}}+4 \left(n^2-1\right) \left[-\left(u_y^2+1\right)\log(u_y) +u_y^2-1\right] u_R+\ocal\left(u_R^2\right) \right\}\,,
\end{multlined}
\end{align}
where we truncated the less important $\delta$ at lower orders to avoid making the equation too long.  Plugging the accessory parameter $\gamma=\gamma_1=-\gamma_2$ into \eqref{snclass}, we obtain the classical R\'enyi entropy in a high temperature expansion:
\begin{multline}\label{eq:thclassical}
S_n= \frac{c(n+1)}{12 n} \left\{\log \sinh ^2(2 \pi  T y) +\text{const}
-\frac{(n^2-1)}{6n^2}\[\cosh (8 \pi  T y)-4 \cosh (4 \pi  T y)\]e^{-4\pi TR} \right.\\
\left.+\frac{(n^2-1)}{6n^2} \[\cosh (4 \pi  T y)+2 \cosh (8 \pi  T y)-\cosh (12 \pi  T y)\]e^{-6\pi TR} +\ocal\left(e^{-8\pi TR}\right) \right\}\,.
\end{multline}
For a general $n>1$, the classical R\'enyi entropy depends on the size $R$ of the spatial circle through the subleading terms in a high temperature expansion.  Upon taking the $n\to1$ limit, the leading order term -- which does not depend on $R$ -- agrees precisely with the entanglement entropy \eqref{eq:storus}, while all subleading terms in the high temperature expansion drop out because they come with additional powers of $n-1$. Thus, similarly to our discussion of the two interval case at the classical level around equation (\ref{eq:tworn}), classical holographic entanglement exhibits a certain simplification in the $n \to 1$ limit, hiding qualitative effects. As we will see below, the one-loop contributions to the $n \to 1$ entanglement entropy do on $R$, as we should generically expect.

Together with \eqref{eq:tgamma} we also find the solutions to the lowest few orders in $u_R$:
\begin{multline}\label{eq:tpsi}
\psi^+(u)=  \frac{1}{\sqrt u} (u-u_y)^{\Delta_+} \(u-\frac{1}{u_y}\)^{\Delta_-} \left\{1+\frac{\left(n^2-1\right) \left(u_y^2-1\right)^2 u_R^2}{24 n^3 u^2 u_y^3}  \times \right.\\
\left.\phantom{\frac{1}{2}}\times \left[u \left((n+1) u^2+n-1\right) u_y^2+u \left((n-1) u^2+n+1\right)-n \left(u^4+1\right) u_y\right] +\ocal\left(u_R^3\right) \right\} \,,
\end{multline}
and the other solution $\psi^-$ may be obtained by applying the map $u \mapsto 1/u$ on $\psi^+$. This map corresponds to $z\mapsto -z$ and is a symmetry of the torus differential equation \eqref{eq:diffeqtorus} with the interval chosen as $[-y,y]$.  Note that the Laurent expansion together with the condition \eqref{eq:uyexpand} ensures that at the $k$-th order in $u_R$, we get a finite combination of different powers of $u$, with powers between $-k$ and $k$.

Now that we have the independent solutions $\psi^{\pm}(u)$, we can find the monodromy matrix (on each sheet) around the spatial circle $z \sim z+R$.  To do this we form two new solutions $\psi^{\pm}(u/u_R)$.  These are the solutions that we get after shifting the old ones by $R$ and wrapping around in the spatial direction.  We then find the monodromy matrix $L_1$ by matching these two sets of solutions:
\beq\label{eq:L1t}
\begin{pmatrix}\psi^+(u/u_R) \\
\psi^-(u/u_R) \\
\end{pmatrix}=L_1 \begin{pmatrix}\psi^+(u) \\
\psi^-(u) \\
\end{pmatrix} \,.
\eeq
We have to match these solutions in a region where their expansions \eqref{eq:uexpand} are simultaneously valid.  As we showed earlier, as long as the monodromy around the time circle is trivial, the Laurent expansion for $\psi^{\pm}(u)$ in \eqref{eq:uexpand} converges within an enlarged annulus $u_R/u_y<|u|<u_y/u_R$.  For the shifted solutions $\psi^{\pm}(u/u_R)$ the region of convergence is $u_R^2/u_y<|u|<u_y$.  This means that both sets of solutions are valid within a smaller annulus $u_R/u_y<|u|<u_y$.  In this region we may expand the entire solutions $\psi^{\pm}$ including the prefactor in \eqref{eq:tpsi}.  We then match the coefficients and find $L_1$ from \eqref{eq:L1t} to any order in $u_R$ that we want.

To the lowest few orders we find the matrix elements of $L_1$ and its inverse as
\begin{align}
(L_1)_{11} &= \frac{n u_y^{1-\frac{1}{n}}}{\left(1-u_y^2\right) \sqrt{u_R}} \left\{ 1-\frac{\left[(n+1) u_y^2+n-1\right]^2}{4 n^2 u_y^2} u_R +\ocal(u_R^2) \right\} \,,\nonumber\\
(L_1)_{12} &= \frac{n u_y}{\left(1-u_y^2\right) \sqrt{u_R}} \left\{ 1-\frac{\left[(n+1) u_y^2+n-1\right] \left[(n-1) u_y^2+n+1\right]}{4 n^2 u_y^2}u_R +\ocal(u_R^2) \right\} \,,\nonumber\\ \label{eq:tL1}
(L_1)_{21} &=-(L_1)_{12} \,,\qquad
(L_1)_{22}=-(L_1)_{11}|_{n\to-n} \,,\qquad
L_1^{-1}=L_1|_{n\to-n} \,.
\end{align}

Finally, we can find all the other Schottky generators $L_i$, $i=2,\cdots,n$ by conjugating $L_1$ with the monodromy matrix $M_2$ around $z_2=y$:
\beq\label{eq:tLi}
L_i=M_2^{i-1} L_1 M_2^{-(i-1)} \,,\qquad
M_2=\begin{pmatrix}e^{2\pi i\Delta_{+}} & 0 \\
0 & e^{2\pi i\Delta_{-}}
\end{pmatrix}.
\eeq

\subsection{High temperature limit: one-loop contribution}

In the previous subsection we have described how to solve the torus differential equation \eqref{eq:diffeqtorus} order by order in a high temperature expansion, imposing the trivial monodromy condition around the time circle.  From these solutions we found the generators $L_i$ of the Schottky group.  In this subsection we will calculate the one-loop determinant contributions to the R\'enyi entropies by forming primitive words from the Schottky generators $L_i$ and computing their eigenvalues.  We will do this order by order in a high temperature expansion, arguing that at each order only a finite number of words contribute, similarly to the situation in the two interval case in section \ref{sec:univ}.

We will work at leading order in the large-$T$ limit, although there is no obstacle (in principle) in applying our method systematically.  From \eqref{eq:tL1} and \eqref{eq:tLi} we find that the Schottky generators and their inverses can be written as
\begin{align}
L_k&= \frac{n u_y}{\left(1-u_y^2\right) \sqrt{u_R}} \left(
\begin{array}{cc}
 u_y^{-1/n} & e^{2\pi ik/n} \\
 -e^{-2\pi ik/n} & -u_y^{1/n}
\end{array}
\right)+\ocal(\sqrt{u_R}) \,,\nonumber\\ \label{eq:Lki}
L_k^{-1}&= \frac{n u_y}{\left(1-u_y^2\right) \sqrt{u_R}} \left(
\begin{array}{cc}
 -u_y^{1/n} & -e^{2\pi ik/n} \\
 e^{-2\pi ik/n} & u_y^{-1/n}
\end{array}
\right)+\ocal(\sqrt{u_R}) \,.
\end{align}

Let us consider an arbitrary word $\gamma$ (not to be confused with the accessory parameter) of length $m$ built from the Schottky generators.  When we talk about the length of a word here and below, we always assume that it cannot be shortened by conjugation within the Schottky group.  From \eqref{eq:Lki} we see that a word of length $m$ is of order $u_R^{-m/2}$, and would generically have a larger eigenvalue $q_\gamma^{-1/2}$ which is of order $u_R^{-m/2}$.  Since the one-loop determinant \eqref{eq:zoneloop} is of order $q_\gamma^2$ for small $q_\gamma$, we conclude that a word of length $m$ contributes of order $u_R^{2m}$ and higher to the one-loop R\'enyi entropies in the large-$T$ limit, assuming that the leading $u_R^{-m/2}$ term in the larger eigenvalue does not vanish.  We now prove this by explicitly computing the leading term.

We note from \eqref{eq:Lki} that $L_k$ and $L_k^{-1}$ are all of the form
\beq
\left(\begin{array}{cc}
 \alpha & \beta \\
 \lambda\alpha & \lambda\beta
\end{array}\right) \,.
\eeq
This form is preserved under multiplication:
\beq
\left(\begin{array}{cc}
 \alpha_1 & \beta_1 \\
 \lambda_1\alpha_1 & \lambda_1\beta_1
\end{array}\right)
\left(\begin{array}{cc}
 \alpha_2 & \beta_2 \\
 \lambda_2\alpha_2 & \lambda_2\beta_2
\end{array}\right)
= (\alpha_1+\lambda_2\beta_1)
\left(\begin{array}{cc}
 \alpha_2 & \beta_2 \\
 \lambda_1\alpha_2 & \lambda_1\beta_2
\end{array}\right) \,.
\eeq
Using this equation repeatedly, we find that any word $\gamma$ of length $m$ may be written as
\begin{multline}
L_{k_1}^{\s_1} L_{k_2}^{\s_2} \cdots L_{k_m}^{\s_m} = \[\frac{n u_y}{\left(1-u_y^2\right) \sqrt{u_R}}\]^m \[\prod_{j=1}^{m-1} \s_j \(u_y^{-\s_j/n} - e^{2\pi i (k_j-k_{j+1})/n} u_y^{\s_{j+1}/n}\)\] \times\\
\times \left(\begin{array}{cc}
 \s_m u_y^{-\s_m/n} & \s_m e^{2\pi i k_m/n} \\
 -\s_m e^{-2\pi i k_1/m} u_y^{(\s_1-\s_m)/n} & -\s_m e^{2\pi i (k_m-k_1)/n} u_y^{\s_1/n}
\end{array}\right) +\ocal\(u_R^{-m/2+1}\) \,,
\end{multline}
where $\sigma_j$, $j=1,2,\cdots,m$ are $\pm1$.  The larger eigenvalue of this word can easily be worked out at leading order:
\beq\label{eq:tqgamma}
q_\gamma^{-1/2} = \[\frac{n u_y}{\left(1-u_y^2\right) \sqrt{u_R}}\]^m \prod_{j=1}^{m} \s_j \(u_y^{-\s_j/n} - e^{2\pi i (k_j-k_{j+1})/n} u_y^{\s_{j+1}/n}\) + \ocal\(u_R^{-m/2+1}\)\,,
\eeq
where $k_{m+1}$ and $\s_{m+1}$ are understood as $k_1$ and $\s_1$.
This leading term cannot vanish; for the $j$-th factor of the product to be zero, we have to demand $k_j=k_{j+1}$ and $\s_j=-\s_{j+1}$, and then we should have contracted the $j$-th letter and the next (which is its inverse), shortening the whole word.  Since we have assumed that this is not the case from the start, we conclude that the leading term in \eqref{eq:tqgamma} does not vanish.

Therefore we have shown that the leading order contributions to the one-loop determinant come from single-letter words $L_k$ and $L_k^{-1}$.  In this sense single-letter words are analogs of the CDWs discussed in section \ref{sec:univ}, two-letter words are analogs of 2-CDWs, and so on.

Let us now calculate the contributions from single-letter words.  From \eqref{eq:tqgamma} we find that the larger eigenvalue of $L_k^\s$ is independent of $k$ or $\s$:
\beq\label{eq:q1}
q_1^{-1/2} = \frac{n u_y (u_y^{-1/n} - u_y^{1/n})}{(1-u_y^2) \sqrt{u_R}} +\ocal(\sqrt{u_R})
= \frac{n \sinh \left(\frac{2 \pi T y}{n}\right)}{\sinh(2 \pi T y)} e^{\pi TR} +\ocal\left(e^{-\pi TR}\right) \,,
\eeq
where the subscript 1 in $q_1$ denotes single-letter words.  We can now easily sum over these words in \eqref{eq:zoneloop} and get the one-loop partition function on the $n$-sheeted cover
\beq\label{eq:thzloop}
\left. \log Z_n \right|_\text{one-loop} = \sum_{\gamma\in\mathcal P} \Real \[q_\gamma^2+\ocal(q_\gamma^3)\] = \frac{2\sinh^4(2 \pi T y)}{n^3 \sinh^4\left(\frac{2 \pi T y}{n}\right)} e^{-4\pi TR} + \ocal\(e^{-6\pi TR}\) \,.
\eeq
Unlike the two interval case, this does not vanish at $n=1$ (which is expected since it is the one-loop determinant of the original bulk space), and we subtract it off in the one-loop R\'enyi entropy as in \eqref{eq:sz} 
\beq\label{eq:thloop}
\left. S_n \right|_\text{one-loop} = -\frac{1}{n-1}\[\frac{2\sinh^4(2 \pi T y)}{n^3 \sinh^4\left(\frac{2 \pi T y}{n}\right)}-2n\] e^{-4\pi TR} + \ocal\(e^{-6\pi TR}\) \,.
\eeq
Finally, we take the $n\to1$ limit and get the one-loop contribution to the entanglement entropy in a high temperature expansion:
\beq
\left. S \right|_\text{one-loop} = \[-16 \pi T y \coth(2 \pi T y)+8\] e^{-4 \pi T R}+ \ocal\(e^{-6\pi TR}\) \,.
\eeq
We note that the one-loop entanglement entropy depends on $R$ (even at leading order).  This may be compared to the classical result (\ref{eq:hightclass}), where the entanglement entropy is strictly independent of $R$ above the Hawking-Page temperature.

\subsection{Low temperature limit}

In this subsection we present the results in the low temperature limit, which is defined as small $T$ in units of $1/R$ with arbitrary $y/R$.  In this limit we may apply the same techniques that we used at high temperatures.  We can take a shortcut by applying the following transformation to the torus differential equation \eqref{eq:diffeqtorus} and its solutions:
\beq\label{eq:transform}
R \to \frac{i}{T} \,, \qquad
T \to -\frac{i}{R} \,.
\eeq
This transforms a high temperature solution (with trivial monodromy around the time circle) to a low temperature solution (with trivial monodromy around the spatial circle).

Applying this transformation to \eqref{eq:thclassical}, we obtain the classical R\'enyi entropy at low temperatures:
\begin{multline}
S_n= \frac{c(n+1)}{12 n} \left\{\log \sin^2\(\frac{2\pi y}{R}\) +\text{const}
-\frac{(n^2-1)}{6n^2}\[\cos\(\frac{8\pi y}{R}\)-4 \cos\(\frac{4\pi y}{R}\)\]e^{-\frac{4\pi}{TR}} \right.\\
\left.+\frac{(n^2-1)}{6n^2} \[\cos\(\frac{4\pi y}{R}\)+2 \cos\(\frac{8\pi y}{R}\)-\cos\(\frac{12\pi y}{R}\)\]e^{-\frac{6\pi}{TR}} +\ocal\left(e^{-\frac{8\pi}{TR}}\right) \right\}\,.
\end{multline}
Analogously to the high temperature case, we note that the subleading terms depend on $T$, but they drop out in the $n\to1$ limit for the entanglement entropy, giving exact agreement with \eqref{eq:lowT}.

Applying the transformation \eqref{eq:transform} to \eqref{eq:thloop} we get the one-loop R\'enyi entropy at low temperatures\footnote{We should in principle be careful about taking the real part of $q_\gamma^2$ in \eqref{eq:thzloop} -- in other words, we should first apply the transformation to $q_\gamma^2$ and then take the real part.  In our case it does not matter because $q_1^{-1/2}$ from \eqref{eq:q1} is real before and after the transformation (at least to leading order).}
\beq
\left. S_n \right|_\text{one-loop} = -\frac{1}{n-1}\[\frac{2\sin^4\(\frac{2 \pi y}{R}\)}{n^3 \sin^4\left(\frac{2 \pi y}{nR}\right)}-2n\] e^{-\frac{4\pi}{TR}} + \ocal\(e^{-\frac{6\pi}{TR}}\) \,,
\eeq
as well as the one-loop entanglement entropy
\beq
\left. S \right|_\text{one-loop} = \[-\frac{16 \pi y}{R} \cot\(\frac{2 \pi y}{R}\)+8\] e^{-\frac{4\pi}{TR}} + \ocal\(e^{-\frac{6\pi}{TR}}\) \,.
\eeq
This result agrees with the asymptotic low temperature form obtained in \cite{Herzog:2012bw} for free theories with a mass $m$, under the identification $4 \pi/R \sim m$. Our result reveals a temperature dependence that is absent classically in \eqref{eq:lowT} below the Hawking-Page transition. Furthermore the one-loop correction exhibits the anticipated difference of the entanglement entropy of the interval and its complement
\be
S\Big([-y,y]\Big) - S\Big([-R/2,R/2] \setminus [-y,y]\Big) = - 8 \pi \cot\(\frac{2 \pi y}{R}\)e^{-\frac{4\pi}{TR}} + \ocal\(e^{-\frac{6\pi}{TR}}\) \,.
\ee
That is, the one-loop correction shows the impureness of the thermal state of the CFT.

The one-loop high and low temperature results we have just obtained are very similar in form to the finite temperature entanglement entropies of free fermions in 1+1 dimensions, computed in \cite{Herzog:2013py}. It seems likely that, in these limits, universal results analogous to the small cross-ratio results (\ref{s1nx}) and (\ref{s11x}) can be obtained.\footnote{This observation arose in discussions with Chris Herzog.}

\section*{Acknowledgements}

It is a pleasure to acknowledge helpful discussions with Tom Faulkner, A. Liam Fitzpatrick, Daniel Harlow, Chris Herzog, Diego Hofman, Nabil Iqbal, Jared Kaplan, Juan Maldacena, and Eva Silverstein. This work is partially funded by a DOE Early Career Award (SAH and VLM), a Sloan fellowship (SAH), NSF graduate fellowship number DGE-114747 (TB), NSF grant number PHY-0756174 (XD) and the Templeton foundation (SAH). XD would like to acknowledge the KITP program on ``Primordial Cosmology'' for hospitality during part of this work. Any opinion, findings, and conclusions or recommendations expressed in this material are those of the authors and do not necessarily reflect the views of the National Science Foundation.

\appendix

\section{Weierstrass functions}
\label{sec:weierstrass}

The Weierstrass elliptic function $\wp$ is defined as
\beq\label{eq:wpdef}
\wp(z) \equiv \wp \(z;R,\frac{i}{T}\) = \frac{1}{z^2} + \sum_{(m,n)\ne(0,0)} \[ \frac{1}{\(z+mR + \frac{in}{T}\)^2} - \frac{1}{\(mR + \frac{in}{T}\)^2}\] \,,
\eeq
where the two periods $\{R,i/T\}$ are chosen to agree with the torus identification \eqref{eq:period}.  We may perform the sum over either $m$ or $n$ in closed form.  For example, summing over $n$ we get
\beq\label{eq:wpsum}
\wp(z) = \sum_{m=-\infty}^{\infty} \frac{\pi^2 T^2}{\sinh^2\[\pi T(z+mR)\]} - \sum_{m\ne0} \frac{\pi^2 T^2}{\sinh^2(\pi mTR)} + \frac{\pi^2 T^2}{3} \,.
\eeq
This formula is particularly useful for the high temperature expansion, as the summand becomes exponentially suppressed as we increase $m$.  If we had chosen to perform the sum over $m$ in \eqref{eq:wpdef}, we would have obtained \eqref{eq:wpsum} with the $R$ and $i/T$ exchanged.

The Weierstrass zeta function $\zeta$ can be obtained from $\wp$ through the relation
\beq
\wp(z) = -\zeta'(z) \,.
\eeq
This relation fixes $\zeta(z)$ up to an integration constant which cancels out from the torus differential equation \eqref{eq:diffeqtorus}. Using the single-sum version \eqref{eq:wpsum} we get
\beq\label{eq:zetasum}
\zeta(z) = \sum_{m=-\infty}^{\infty} \pi T \coth\[\pi T(z+mR)\]+ \sum_{m\ne0} \frac{\pi^2 T^2 z}{\sinh^2(\pi mTR)} - \frac{\pi^2 T^2 z}{3} +\text{const} \,.
\eeq
We can also recall that the Weierstrass sigma function $\sigma$ is defined by
\be
\zeta(z) = \frac{\sigma'(z)}{\sigma(z)} \,.
\ee

\section{The 2-CDW contribution to the R\'enyi entropies}
\label{sec:2cdw}

In this appendix we obtain a closed form for the sum
\beq\label{hn}
h(n) \equiv \sum_{\{m_j\}} \frac{1}{\sin^4\(\frac{\pi(m_1-m_2)}{n}\) \sin^4\(\frac{\pi(m_2-m_3)}{n}\) \sin^4\(\frac{\pi(m_3-m_4)}{n}\) \sin^4\(\frac{\pi(m_4-m_1)}{n}\)} \,,
\eeq
for integer $n\ge 1$. The range of the sum was given in \eqref{sn2r}.  We will use a generalization of the `residue' method of  Appendix A.2 of \cite{Headrick:2010zt}.

Let us first do the following sum
\beq\label{ga}
g_{\alpha}(n;m_1,m_3)=\sum_{m=0,\, m\ne m_1,m_3}^{n-1} \frac{1}{\sin^{2\alpha}\(\frac{\pi(m_1-m)}{n}\) \sin^{2\alpha}\(\frac{\pi(m-m_3)}{n}\)},
\eeq
where $\alpha$ is a positive integer, and $m_1$, $m_3$ are integers between 0 and $n-1$ (inclusive).  In order to perform this sum in closed form, we consider the following meromorphic function
\beq\label{eq:faz}
f_\alpha(z)=2^{4\alpha}\frac{n}{z(z^n-1) (z-z_1)^\alpha (z^{-1}-z_1^{-1})^\alpha (z-z_3)^\alpha (z^{-1}-z_3^{-1})^\alpha},
\eeq
where $z_1\equiv e^{2\pi i m_1/n}$, $z_3\equiv e^{2\pi i m_3/n}$.  This meromorphic function has poles only at the $n$-th roots of unity.  It has no branch cuts because $\alpha$ is a positive integer, and the behavior at infinity is regular.  Therefore the sum of residues at the $n$ poles must vanish.  At any $n$-th root of unity that is not equal to $z_1$ or $z_3$, the residue is
\beq
\text{Res}(f_\alpha,e^{2\pi i m/n})= \frac{1}{\sin^{2\alpha}\(\frac{\pi(m_1-m)}{n}\) \sin^{2\alpha}\(\frac{\pi(m-m_3)}{n}\)}, \quad
m\ne m_1,m_3.
\eeq
This is precisely the summand in \eqref{ga}.  Therefore we arrive at
\beq\label{ga2}
g_\alpha(n;m_1,m_3)=-\sum_{m\in\{m_1,m_3\}}\text{Res}(f_\alpha,e^{2\pi im/n}).
\eeq

We now need to distinguish between two cases: $m_1=m_3$ and $m_1\ne m_3$.  When $m_1=m_3$, we can actually allow $\alpha$ to be either an integer or a half-integer without introducing branch cuts in (\ref{eq:faz}), and we find that $g_\alpha(n;m_1,m_1)$ is a polynomial in $n$ of degree $4\alpha$ and is independent of $m_1$, so we can abbreviate it as $g_\alpha(n)$:
\beq\label{gan}
g_\alpha(n)\equiv g_\alpha(n;m_1,m_1)= -\text{Res}(f_\alpha|_{m_3=m_1},e^{2\pi im_1/n}).
\eeq
For a given positive integer $\alpha$ this can be worked out explicitly; for example, for $\a=1$ and $2$ we get
\begin{align}
g_1(n) &= \frac{1}{45} \left(n^4+10 n^2-11\right),\\
g_2(n) &= \frac{3 n^8+40 n^6+294 n^4+2160 n^2-2497}{14175}.
\end{align}
Upon taking the $n\to1$ limit, we find that
\beq
\lim_{n\to1} \frac{g_\alpha(n)}{n-1} = \frac{\Gamma(3/2)\Gamma(2\alpha+1)} {\Gamma(2\alpha+3/2)},
\eeq
in full agreement with (and deriving for $2\alpha\in \mathbb Z^+$) \eqref{eq:continue}.
When $m_1\ne m_3$, we evaluate \eqref{ga2} for $\alpha=2$ and get
\bea
\lefteqn{g_2(n;m_1,m_3)=} \\
&& \frac{300 \left(n^2+11\right) \sin^2 \left(\frac{\pi(m_1-m_3)}{n}\right)+2 \left(n^4-110 n^2-251\right) \sin^4 \left(\frac{\pi(m_1-m_3)}{n}\right)-3150}{45 \sin^8 \left(\frac{\pi(m_1-m_3)}{n}\right)}. \nonumber
\eea

We are now ready to do the sum over $m_2$ and $m_4$ in \eqref{hn}.  For $m_1\ne m_3$, we simply get
\beq\label{g22}
g_2(n;m_1,m_3)^2= \frac{4 \csc ^8 \left(n^4-110 n^2+150 \left(n^2+11\right) \csc ^2-251-1575 \csc ^4\right)^2}{2025},
\eeq
where $\csc\equiv\csc\(\pi(m_1-m_3)/n\)$.  We may now sum over $m_1\ne m_3$ and then over $m_3$, by replacing each $\csc^{4\alpha}$ term in the polynomial expansion of \eqref{g22} (as a polynomial in $\csc$) with $n g_\alpha(n)$ which can be worked out from \eqref{gan}.

We combine this with the contribution from $m_1=m_3$.  Here we need to impose $m_2\ne m_4$.  We may first do the sum without imposing this condition, getting $n g_2(n)^2$, and then subtract the contribution from $m_2=m_4$ which is $n g_4(n)$.  Here the factor of $n$ comes from the final trivial sum over $m_1=m_3$.

Combing the contributions from $m_1\ne m_3$ and $m_1=m_3$, we finally arrive at the sum \eqref{hn} in closed form:
\begin{multline}
h(n)=\frac{4 (n-2) (n-1) n (n+1) (n+2)}{488462349375}\left(5703 n^{12}+192735 n^{10}+3812146 n^8\right.\\
\left.+75493430 n^6+1249638099 n^4+9895897835 n^2-162763727948\right).
\end{multline}

\end{document}